\documentclass[twocolumn]{aastex631}

\usepackage{threeparttable}
\usepackage{parskip}
\usepackage{float}
\usepackage{hyperref}
\usepackage{graphicx}
\usepackage{booktabs}
\usepackage{amsthm}
\usepackage{amsmath}
\usepackage{soul}
\usepackage{xspace}

\newcommand{\fermi}{{\it Fermi\/}\xspace}

\newcommand{\fourFGL}{4FGL\xspace}
\newcommand{\fsrq}{FSRQ\xspace}
\newcommand{\fsrqs}{FSRQs\xspace}
\newcommand{\bll}{BL~Lac\xspace}
\newcommand{\blls}{BL~Lacs\xspace}
\newcommand{\bcu}{BCU\xspace}
\newcommand{\bcus}{BCUs\xspace}

\shorttitle{PU Learning for Fermi BCU Classification}
\shortauthors{Akbar et al.}

\begin{document}

\title{Classification of \fermi~Blazars of Uncertain Type Using
       Positive-Unlabelled Machine Learning}

\author[0009-0005-3498-1104]{Sikandar Akbar}
\affiliation{Department of Physics, University of Kashmir, Srinagar 190006, India}
\email{darprince46@gmail.com}

\author[0000-0003-1458-4396]{Zahir Shah}
\affiliation{Manipal Centre for Natural Sciences, Centre of
Excellence, Manipal Academy of Higher Education, Manipal 576104, India}
\email{zahir.shah@manipal.edu}

\author[0000-0002-8955-3212]{Zahoor Malik}
\affiliation{Department of Physics, University of the Free State, Bloemfontein 9301, South Africa}

\author{Athar A. Dar}
\affiliation{Department of Physics, University of Kashmir, Srinagar 190006, India}

\begin{abstract}

We apply positive-unlabelled (PU) machine learning to the 1623 blazars
of uncertain type (\bcus) in the \fermi~Large Area Telescope Fourth Source Catalogue, Data Release 4 (\fourFGL-DR4). Five PU classifiers are trained on seven gamma-ray features, once with the 820 flat-spectrum radio quasars (\fsrqs) and once with the 1490 BL~Lac objects (\blls) as the labelled class (FSRQ mode and BL~Lac mode). Each source is scored out of fold over
100 repeated five-fold splits, and its confidence is the fraction of
splits in which at least three of the five classifiers flag it. At a
confidence of at least 0.1, 437 \bcus~are FSRQ-like, 597 are BL~Lac-like,
12 are flagged in both modes and 577 in neither. On known sources the ensemble reaches a recall of 0.80 in \fsrq{} mode and 0.82 in \bll{} mode at a false-positive rate of 0.08, but a logistic regression trained on the labelled sources alone
does at least as well, so PU learning gives no accuracy gain here.
Removing the sky coordinates from the features does not lower the performance
on known sources but adds 107 \bcus~to the candidate lists and drops 30,
which is consistent with the coordinates acting as a selection shortcut,
so we exclude them. We check the candidates against published optical spectra, Sloan Digital Sky Survey V (SDSS-V) spectra, and \bcus~that later received a class in the
catalogue. Among sources assigned a single class, BL~Lac-like calls are
correct for 84 to 100\% of the sources and FSRQ-like calls for 58 to
83\%, depending on the sample. These samples come from spectroscopic
follow-up and do not represent all \bcus. The candidates are fainter and
lie closer to the Galactic plane than the labelled sources, and the
excess of unclassified \bcus~in the plane is not explained. We release
the confidences of all 1623 \bcus.

\end{abstract}


\section{Introduction}
\label{sec:intro}

Blazars are active galactic nuclei (AGN) with a relativistic jet
pointed close to the line of sight \citep{1995PASP..107..803U}. They
are the largest population of identified extragalactic sources in the
\fermi~Large Area Telescope (LAT) catalogues
\citep{2009ApJ...697.1071A,2020ApJS..247...33A}. Blazars are split by
their optical spectra into flat-spectrum radio quasars (\fsrqs), with
broad emission lines, and BL~Lac objects (\blls), whose lines are weak
or absent (equivalent width below 5~\AA). \fsrqs~are luminous and have
radiatively efficient accretion discs, and their gamma-ray emission is
dominated by inverse Compton scattering of external radiation
\citep{2010MNRAS.402..497G}. \blls~are less luminous and are mostly
synchrotron self-Compton emitters \citep{2010MNRAS.402..497G}. Together
they form the blazar sequence, in which the synchrotron peak frequency
falls as luminosity rises
\citep{1998MNRAS.299..433F,2017MNRAS.469..255G}. Whether the two
classes are physically distinct or joined by transition objects is
debated \citep{2011MNRAS.414.2674G}.

The \fermi~\fourFGL~\citep{2020ApJS..247...33A,2023arXiv230712546B}
contains 3933 blazars with a class label: 820 \fsrqs, 1490 \blls, and
1623 blazars of uncertain type (\bcus). A \bcu~has gamma-ray properties
consistent with a blazar but no conclusive optical spectral
classification. The spectroscopic coverage of known blazars is not
random. \citet{2023ApJ...946..109A} notes that known \blls~are
concentrated in the northern Galactic hemisphere because southern
sources have less optical follow-up. In our sample, \bcus~also differ
from \fsrqs~and \blls~in Galactic latitude, gamma-ray flux and
catalogue age (Section~\ref{sec:data}). Assigning \bcus~to a subclass
matters for the cosmic evolution of \blls~\citep{2014ApJ...780...73A},
for the contribution of blazars to the diffuse gamma-ray background
\citep{2018ApJ...856..106D}, and for tests of blazar counterparts of
astrophysical neutrinos \citep{2016MNRAS.457.3582P}.

Many studies have classified \bcus~and other \fermi~sources with
supervised machine learning
\citep{2012ApJ...753...83A,2012MNRAS.424L..64M,2013MNRAS.428..220H,
2014ApJ...782...41D,2016MNRAS.462.3180C,2017A&A...602A..86L,
2017MNRAS.470.1291S,2019MNRAS.490.4770K,2020MNRAS.493.1926K,
2019ApJ...887..134K,2022Univ....8..436F,2023ApJ...946..109A}.
These studies use random forests, support vector machines, neural
networks and gradient boosted trees. The most comparable study,
\citet{2023ApJ...946..109A}, trained five classifiers on the known
\fsrqs~and \blls~of the Clean sample of the Fourth Catalog of AGN detected by the \fermi-LAT, Data Release 3 (4LAC-DR3) (670 \fsrqs, 1335
\blls) and applied them to 1115 \bcus. A \bcu~was labelled only if all
five classifiers agreed, which gave 610 \blls~and 333 \fsrqs. The
combined classifier had an accuracy of $0.938\pm0.012$ and an area under the receiver operating characteristic (ROC) curve (AUC) of $0.964\pm0.014$ on known sources. Two limits follow from this design.
Every \bcu~must be put in one of two classes, or left unlabelled when
the votes disagree. And the accuracy is measured on known sources,
which differ from \bcus~in the ways described above, so it does not
measure accuracy on \bcus.
Recent studies use smaller samples. \citet{Afroz2025} used 988 known
blazars (790 for training) to train a neural network and classified 82 \bcus, and
\citet{Chen2026} classified 80 of 86 \bcus~with SDSS spectra by the
equivalent width of their emission lines.

Positive-unlabelled (PU) learning offers a different framing
\citep{2020MLear.109..719B}. One class, for example \fsrqs, is the
positive class with known labels. All other sources, including
\bcus~and the other known class, are unlabelled: each may or may not
belong to the positive class. Under the Selected Completely At Random
(SCAR) assumption, in which labelled positives are a random sample of
all positives, the posterior probability of being positive is
$P(y=1|\mathbf{x}) = P(s=1|\mathbf{x})/c$, where $s$ is the label
indicator and $c$ is the fraction of positives that are labelled
\citep{2008KDD....14..213E}. We run the method twice, with \fsrqs~and
then \blls~as the positive class. Each run scores every \bcu~for
similarity to one known class without requiring it to belong to the
other. A \bcu~can therefore be flagged in both runs or in neither,
which a two-class classifier cannot report. This framing does not use
the other known class as negatives. We test whether that costs
accuracy by comparing with a supervised baseline trained on the
labelled sources only (Section~\ref{sec:baseline}).
SCAR is an idealisation here, because spectroscopically confirmed
sources are not a uniform sample of their parent populations. We test how much the candidate lists depend on the sky coordinates
(Section~\ref{sec:ablation}).

\citet{2024MNRAS.533.3283S} applied PU learning to fast radio burst
(FRB) repeaters. Known repeaters were the labelled positives, and
apparent non-repeaters were unlabelled and may contain hidden
repeaters. Their ensemble of five PU classifiers identified 66
repeater candidates, 18 of which were not found by earlier
machine-learning studies of the same catalogue. The problem has the
same structure as \bcu~classification. We use the same five
classifiers and the same candidate rule, with changes described in
Section~\ref{sec:method}. To our knowledge, this is the first
application of PU learning to blazar subclassification.

We run five PU classifiers, tuned with the Tree-structured Parzen
Estimator in Optuna \citep{2019KDD....14.2623A}, in the two modes
described above. A \bcu~flagged in both modes is not clearly assigned
to either class by its gamma-ray features. Such sources may be
transition objects between the two classes, or sources whose features
are too poorly measured to separate them. We list them as targets for
multiwavelength follow-up. We do not validate the candidate lists with
other models fitted to the same features, because agreement between
such models shows that the features separate the known classes, not that
the \bcu~labels are correct. Instead, we compare the lists with
classifiers trained on the labelled sources only, test how much they
depend on the sky coordinates, and check them against published optical
spectra and against \bcus~that were later classified in the catalogue
(Section~\ref{sec:external}). We provide a table of confidences for all
1623 \bcus.

The paper is organised as follows. Section~\ref{sec:data} describes the
data and feature selection. Section~\ref{sec:method} describes the
methodology. Section~\ref{sec:results} presents the results.
Section~\ref{sec:discussion} discusses them, and
Section~\ref{sec:conclusions} gives our conclusions.
\section{Data}
\label{sec:data}

\subsection{The Fermi 4FGL Catalogue}

We use the \fermi~\fourFGL~Data Release 4
\citep[4FGL-DR4;][]{2023arXiv230712546B,2022ApJS..260...53A},
accessed via the FITS file \texttt{gll\_psc\_v35\_f.fit}
(catalogue version v35). The catalogue contains 7195 sources.
We select the 3933 blazars with \texttt{CLASS1} equal to
\texttt{FSRQ}, \texttt{BLL} or \texttt{BCU} (upper or lower case):
820~\fsrqs, 1490~\blls, and 1623~\bcus. Only 66 of the 2310 labelled
sources are firm identifications (upper case); the others are
associations. No cut on the catalogue analysis flags is applied. The
flags are set for 28\% of the \bcus, 11\% of the \blls and 20\% of the
\fsrqs.

\subsection{Feature Selection}
\label{sec:features}

We use seven gamma-ray features from the catalogue
(Table~\ref{tab:features}). They describe the spectral shape,
spectral curvature, brightness and variability of each source, and
all are available for the full sample except \texttt{log\_Flux\_Peak}
(see below) and \texttt{LP\_beta}, which is missing for one source. Five other catalogue columns were not used because each
is strongly correlated with a retained column:
\texttt{LP\_Index} ($|\rho| = 0.97$ with \texttt{PL\_Index}),
\texttt{Energy\_Flux100} ($|\rho| = 0.95$ with
\texttt{log\_Flux1000}), \texttt{Frac\_Variability}
($|\rho| = 0.88$ with \texttt{log\_Variability\_Index}),
\texttt{Signif\_Avg} ($|\rho| = 0.89$ with \texttt{log\_Flux1000}),
and \texttt{log\_LP\_EPeak} ($|\rho| = 0.81$ with
\texttt{log\_Pivot\_Energy}).

\begin{figure}[ht]
\centering
\includegraphics[width=\columnwidth]{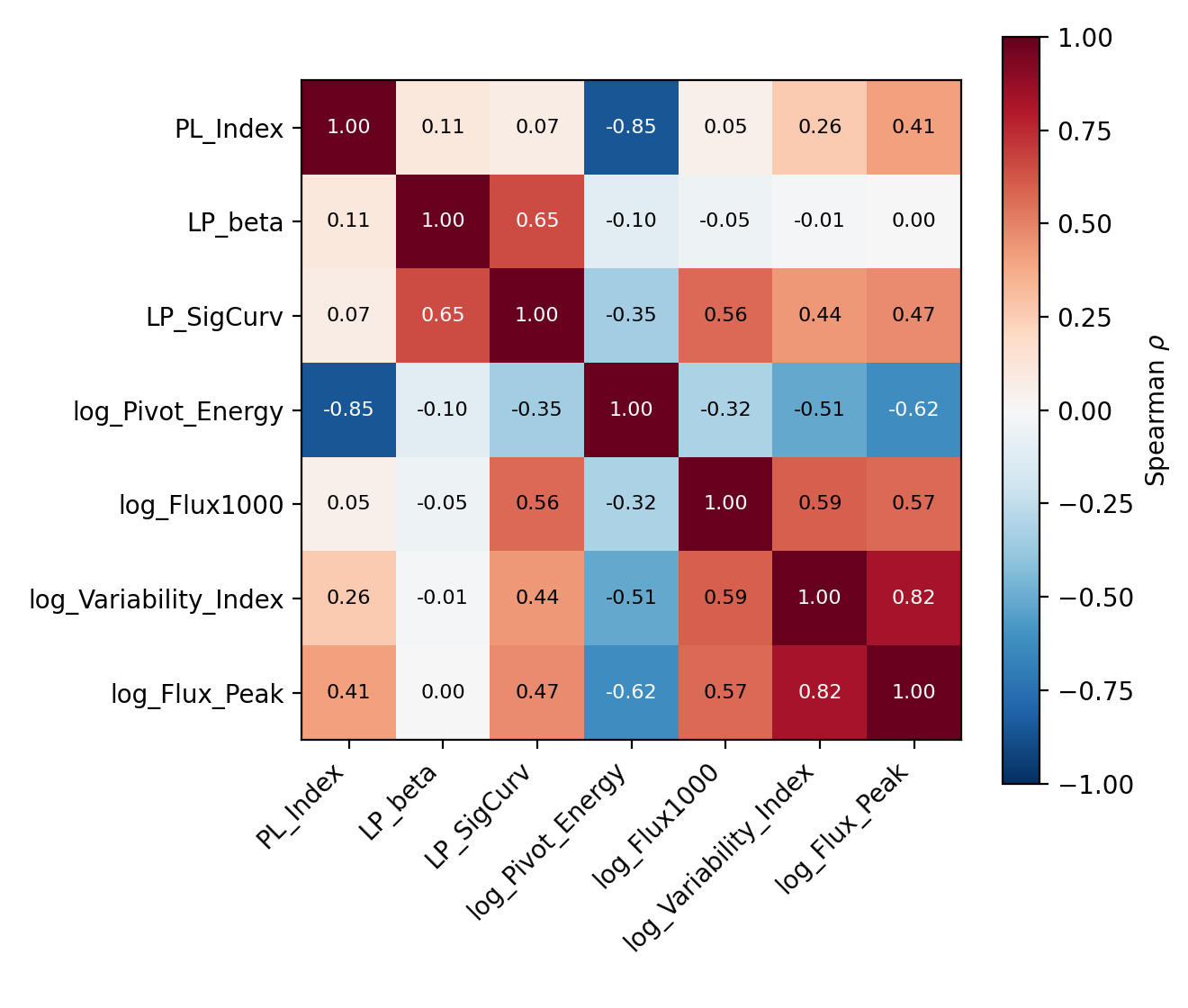}
\caption{
  Spearman rank correlation matrix of the seven features, computed
  pairwise on the non-imputed values. The strongest correlation is
  between \texttt{PL\_Index} and \texttt{log\_Pivot\_Energy}
  ($\rho = -0.853$), and the second strongest is between
  \texttt{log\_Variability\_Index} and \texttt{log\_Flux\_Peak}
  ($\rho = 0.825$). Both pairs are kept (Section~\ref{sec:features}). Each cell gives the coefficient $\rho$ for a pair of features, from $-1$ (blue) to $+1$ (red). }
\label{fig:correlation}
\end{figure}

Figure~\ref{fig:correlation} shows the Spearman rank correlation
matrix of the seven features. Two pairs remain strongly correlated:
\texttt{PL\_Index} and \texttt{log\_Pivot\_Energy}
($\rho = -0.853$), and \texttt{log\_Variability\_Index} and
\texttt{log\_Flux\_Peak} ($\rho = 0.825$). We keep both pairs. The correlations split credit between features in the SHapley Additive exPlanations (SHAP) analysis
(Section~\ref{sec:shap_results}).

Four features with heavily skewed distributions
(\texttt{Pivot\_Energy}, \texttt{Flux1000},
\texttt{Variability\_Index} and \texttt{Flux\_Peak}) are
$\log_{10}$-transformed. \texttt{LP\_beta} is missing for one source.
\texttt{Flux\_Peak} is missing for 2271 sources (57.7\%): it is
missing for all sources with \texttt{Variability\_Index} $\leq 27.69$
and present for all 1662 sources above this value. Missing values of
both features are imputed with the median of the training data
(Section~\ref{sec:pu_framework}).
Because the missingness is fixed by the variability index, the imputed
\texttt{log\_Flux\_Peak} also marks whether a source is variable. The
fraction of sources with a value is 75\% for \fsrqs, 38\% for \blls
and 29\% for \bcus, so the missingness alone separates \fsrqs from
\blls with a balanced accuracy of 0.68. We keep the feature and did not test
its removal.

\begin{figure*}[ht]
\centering
\includegraphics[width=0.95\textwidth]{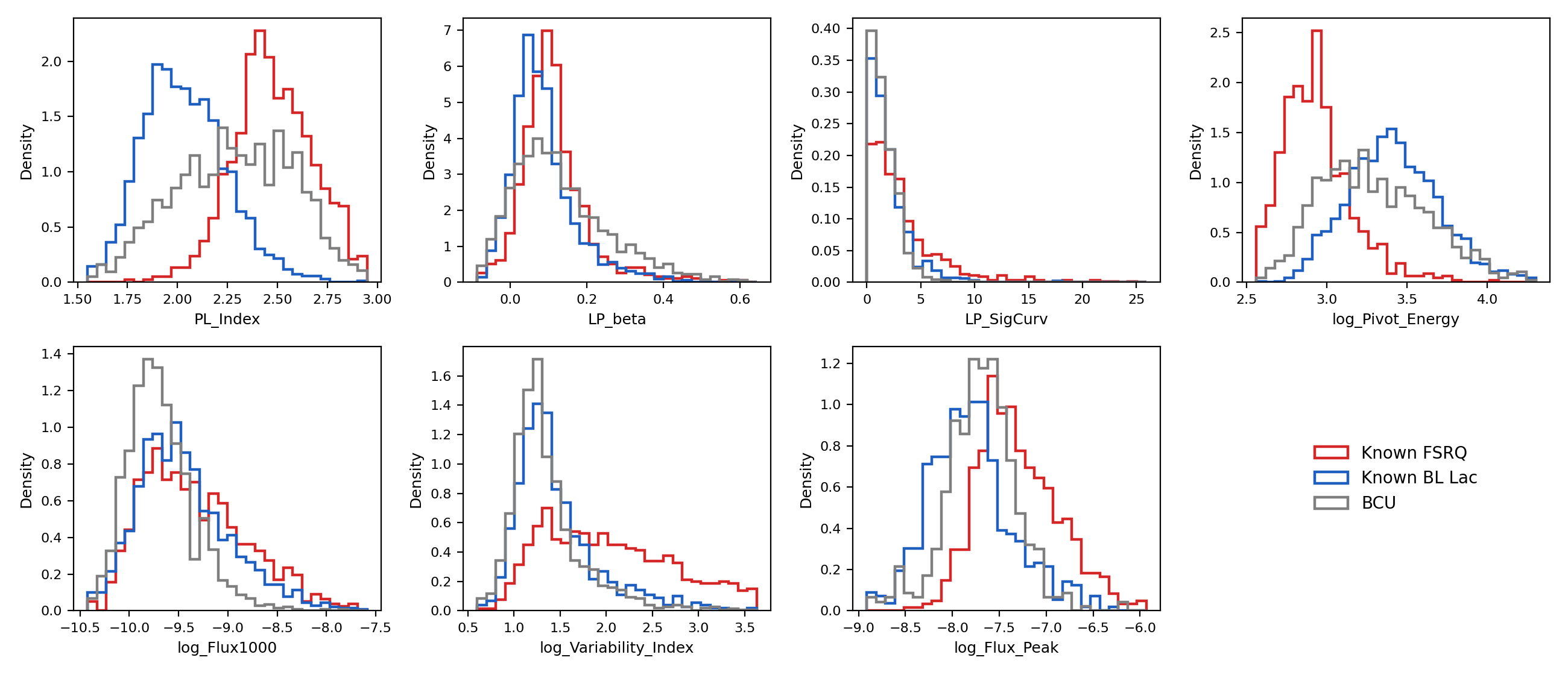}
\caption{
Distribution of the seven gamma-ray features (one histogram per feature, normalised to unit area) for known \fsrqs (red), known \blls (blue), and \bcus (grey), before imputation.
\texttt{log\_Flux\_Peak} is shown only for the 1662 sources for which
it is defined. The power-law spectral index and pivot energy separate
the two confirmed classes best.
}
\label{fig:feature_dist}
\end{figure*}

The \bcus differ from the labelled sources in several ways
(Figure~\ref{fig:feature_dist}). The median \texttt{PL\_Index} is 2.47
for \fsrqs, 2.02 for \blls, and 2.30 for \bcus, and the median
$\log E_\mathrm{pivot}$ is 2.92, 3.40 and 3.26, respectively, so the
\bcus lie between the two classes in both features. They are fainter:
the median integral flux between 1 and 100\,GeV is lower than that of
\blls by a factor of 1.7 and than that of \fsrqs by a factor of 2.1.
They lie closer to the Galactic plane: 18.5\% of the \bcus (300
sources) have $|b| < 10^\circ$, against 4.5\% of the \fsrqs and 5.6\%
of the \blls, and the median $|b|$ is $27.3^\circ$ for the \bcus and
$39.6^\circ$ for both labelled classes. They are also newer: 33\% of
the \bcus entered the catalogue after the first data release, against
14\% of the \fsrqs and 10\% of the \blls. The labelled sources are
therefore not a representative sample of the \bcus.

Galactic coordinates were part of an earlier nine-feature version and
are excluded from the final set (Section~\ref{sec:ablation}). Anderson-Darling (AD) $k$-sample tests between the known \fsrqs and \blls give AD $\geq 17.8$ and
$p \leq 0.001$ for all the physical features, with \texttt{PL\_Index}
and \texttt{log\_Pivot\_Energy} showing the strongest separation
(AD $= 814.5$ and $746.1$). The test for \texttt{log\_Flux\_Peak} uses
only the 617 \fsrqs and 572 \blls for which it is defined. Galactic
longitude shows no significant difference between the classes
(AD $= -0.007$, $p \geq 0.25$), and Galactic latitude a weak one
(AD $= 4.1$, $p = 0.008$). The $p$-values are limited to the range
0.001--0.25 by the implementation. The latitude difference fits the
northern-hemisphere bias in optical spectroscopic follow-up of
\bll~candidates \citep{2023ApJ...946..109A}.

\begin{deluxetable*}{lll}
\tablecaption{Gamma-ray features used in this analysis. Feature is the \fourFGL-DR4 catalogue column name (a \texttt{log\_} prefix marks a $\log_{10}$ transform of the column), Description is the quantity, and Transform is the operation applied before training.\label{tab:features}}
\tablewidth{0pt}
\tablehead{
  \colhead{Feature} &
  \colhead{Description} &
  \colhead{Transform}
}
\startdata
PL\_Index           & Power-law (PL) photon spectral index                      & none        \\
LP\_beta            & LogParabola (LP) spectral curvature $\beta$                 & none        \\
LP\_SigCurv         & Significance of spectral curvature           & none        \\
log\_Pivot\_Energy  & Pivot energy (MeV)                           & $\log_{10}$ \\
log\_Flux1000       & Integral photon flux, 1--100\,GeV            & $\log_{10}$ \\
log\_Variability\_Index & Variability index                       & $\log_{10}$ \\
log\_Flux\_Peak     & Peak flux (ph\,cm$^{-2}$\,s$^{-1}$)          & $\log_{10}$ \\
\enddata
\tablecomments{
\texttt{log\_Flux\_Peak} is defined only for sources with
\texttt{Variability\_Index} $> 27.69$ and is missing for 57.7\% of
the sources (25\% of the \fsrqs, 62\% of the \blls, 71\% of the
\bcus). \texttt{LP\_beta} is missing for one source. Missing values
are imputed with the feature median.
}
\end{deluxetable*}

\section{Methodology}
\label{sec:method}

\subsection{Positive-Unlabelled Learning Framework}
\label{sec:pu_framework}

A supervised classifier trained on known \fsrqs and \blls assigns every
\bcu to one of those two classes, whether or not it belongs to either.
Positive-unlabelled (PU) learning instead takes one known class as the
labelled positive class and treats every other source, including the
other known class, as unlabelled. No source is assigned to a negative
class during training \citep{2020MLear.109..719B}.

Let $\mathbf{x}$ be the feature vector of a source and $s$ its observed
label: $s=1$ for a known member of the positive class and $s=0$ for an
unlabelled source. The true class $y$ is unknown when $s=0$. The class
prior is $\alpha = P(y=1)$ and the label frequency is
$c = P(s=1\,|\,y=1)$. We assume that labelled positives are selected
completely at random (SCAR) from all positives
\citep{2020MLear.109..719B}, that is
\begin{equation}
  P(s=1\,|\,y=1,\mathbf{x}) = P(s=1\,|\,y=1) = c .
  \label{eq:scar}
\end{equation}
It follows that $P(s=1)=c\,\alpha$ and
$P(s=1\,|\,\mathbf{x}) = c\,P(y=1\,|\,\mathbf{x})$, the relation used by
the classifiers below.

SCAR is an idealisation here. Spectroscopic confirmation depends on
brightness and sky position \citep{2023ApJ...946..109A}, so the known
sources are not a uniform sample of their parent population.
Section~\ref{sec:method_validation} tests how far our conclusions depend
on this.

We use the seven features described in Section~\ref{sec:data}, with
missing values replaced by the training-set median and all features
standardised. Both steps are fitted on training data only. We run two
modes:
\begin{itemize}
  \item \textbf{\fsrq{} mode}: the 820 known \fsrqs are positive; the
        remaining 3113 sources (1490 \blls and 1623 \bcus) are unlabelled.
  \item \textbf{\bll{} mode}: the 1490 known \blls are positive; the
        remaining 2443 sources (820 \fsrqs and 1623 \bcus) are unlabelled.
\end{itemize}

\subsection{Classifier Ensemble}
\label{sec:classifiers}

We use the five PU classifiers of \citet{2024MNRAS.533.3283S}: Classic Elkanoto (CE), Weighted Elkanoto (WE), Bagging PU (BC), Modified Logistic Regression (MLR) and PU Extra Trees (PUET). CE and BC are taken from \textsc{pulearn} version 0.0.9 \citep{pulearn}, and WE
is our own implementation (below).
MLR and PUET use the forks of the original authors' code distributed
with \citet{2024MNRAS.533.3283S} as the packages
\texttt{MLR\_ArjunS07\_pip} and \texttt{PUExtraTrees\_ArjunS07\_pip}
(version 0.0.1). The forks change only the class interface and the
project structure; their authors state that the algorithms are those of
\citet{Jaskie2019mlr} and \citet{Wilton2022}. Base classifiers come from
\textsc{scikit-learn} \citep{2011JMLR...12.2825P}.

\textbf{Classic Elkanoto (CE).} A base classifier is trained to predict
$g(\mathbf{x}) \approx P(s=1\,|\,\mathbf{x})$. A fraction of the sample,
the hold-out ratio, is drawn at random from all sources and excluded
from training. The label frequency is estimated as the mean of $g$ over
the labelled positives $H_+$ in that hold-out set,
\begin{equation}
  \hat{c} = \frac{1}{|H_+|}\sum_{\mathbf{x}\in H_+} g(\mathbf{x}).
  \label{eq:ce_c}
\end{equation}
The posterior is
\begin{equation}
  P(y=1\,|\,\mathbf{x}) = \frac{g(\mathbf{x})}{\hat{c}},
  \label{eq:ce_posterior}
\end{equation}
and a source is classified as positive if this exceeds 0.5
\citep{2008KDD....14..213E}. The estimate can exceed 1 and is used only
through the threshold. CE needs $g$ to be well calibrated, so base
classifiers other than logistic regression (LR) and linear discriminant
analysis (LDA) are Platt scaled \citep{Platt1999} with 5-fold
cross-validation. \citet{2024MNRAS.533.3283S} use 10 folds.

\textbf{Weighted Elkanoto (WE).} WE uses the same $g$ and $\hat{c}$ as
CE, with the same calibration. For an unlabelled source, the probability
of being positive is
\begin{equation}
  w(\mathbf{x}) = P(y=1\,|\,\mathbf{x},s=0)
  = \frac{1-\hat{c}}{\hat{c}}\,
    \frac{g(\mathbf{x})}{1-g(\mathbf{x})}
  \label{eq:we_weight}
\end{equation}
\citep{2008KDD....14..213E}. The \textsc{pulearn} implementation
multiplies $g$ by a factor that is computed from the sources passed to it
at prediction time, so the score of a source changes with the batch being
scored. We therefore use our own implementation of the weighted classifier
of \citet{2008KDD....14..213E}, which does not have this property. After
$g$ and $\hat{c}$ are estimated on the hold-out split, each labelled source
enters the training set once as a positive, and each unlabelled source
enters twice: as a positive with weight $w(\mathbf{x})$ and as a negative
with weight $1-w(\mathbf{x})$, with $w$ clipped to $[0,1]$. The base
classifier is refitted on this weighted set, and its predicted probability
is the WE score. For a base classifier that does not accept sample weights
(LDA), the weights are converted to integer row multiplicities (weight
times 20, rounded). The same class is used in tuning and in the
out-of-fold runs.

\textbf{Bagging PU Classifier (BC).} Each of $B$ base classifiers is
trained on the labelled positives and a bootstrap sample of the
unlabelled sources, which act as surrogate negatives, using a fraction
\texttt{max\_samples} of the sources and a fraction \texttt{max\_features}
of the features. A different unlabelled sample is used each time, which
limits the effect of hidden positives in any one sample. The score of a
source is the average over all base classifiers
\citep{2014PaReL..37..201M}. The base classifiers are not calibrated.

\textbf{Modified Logistic Regression (MLR).} MLR fits a logistic model
whose output is bounded above by $\hat{c}$,
\begin{equation}
  g_\text{MLR}(\mathbf{x}) =
  \frac{1}{1 + b^2 + e^{-\mathbf{w}\cdot\mathbf{x}}},
  \label{eq:mlr}
\end{equation}
with $b$ learned together with $\mathbf{w}$, so that
$\hat{c} = 1/(1+b^2)$ \citep{Jaskie2019mlr}. The posterior is
$P(y=1\,|\,\mathbf{x}) = g_\text{MLR}(\mathbf{x})/\hat{c}$. The model is
trained for 250 epochs with a tuned learning rate.
\citet{2024MNRAS.533.3283S} use 100 epochs.

\textbf{PU Extra Trees (PUET).} Extra Trees whose splits minimise a PU
risk estimator \citep{Wilton2022}. For a loss $\ell$ and class prior
$\alpha$, the unbiased estimator (uPU) \citep{2015ICML...37.1386P} is
\begin{equation}
  \hat{R}_\text{uPU} = \alpha\,\hat{\mathbb{E}}_{P}[\ell(f,+1)]
  + \hat{\mathbb{E}}_{U}[\ell(f,-1)]
  - \alpha\,\hat{\mathbb{E}}_{P}[\ell(f,-1)],
  \label{eq:upu}
\end{equation}
and the non-negative estimator (nnPU) \citep{2017NIPS..30.1753K} clips
the last two terms at zero,
\begin{equation}
  \begin{split}
  \hat{R}_\text{nnPU} = {}& \alpha\,\hat{\mathbb{E}}_{P}[\ell(f,+1)] \\
  & + \max\!\bigl\{0,\;
  \hat{\mathbb{E}}_{U}[\ell(f,-1)]
  - \alpha\,\hat{\mathbb{E}}_{P}[\ell(f,-1)]\bigr\}.
  \end{split}
  \label{eq:nnpu}
\end{equation}
Here $P$ and $U$ are the labelled and unlabelled samples. The risk
estimator, the loss (quadratic or logistic), the number of trees (25 to
200), the minimum leaf size (1 to 10), the number of features per split
(square root of the number of features, or all) and
\texttt{max\_candidates} (1 to 10) are tuned. The prior $\alpha$ is
estimated separately for each training set. We fit CE with an LDA base
classifier and a hold-out ratio of 0.3 to obtain $\hat{c}$, and set
$\hat{\alpha} = \min\{P(s=1)/\hat{c},\,0.99\}$ from
equation~(\ref{eq:scar}).

CE, WE and BC each take a tunable base classifier from the three
supervised classifiers with the highest mean recall in that mode
(Appendix~\ref{app:supervised}): AdaBoost, LDA and LR in \fsrq{} mode, and
LDA, LR and a support vector classifier (SVC) in \bll{} mode.

\subsection{Hyperparameter Optimisation}
\label{sec:hyperopt}

Hyperparameters are tuned with the Tree-structured Parzen Estimator in
Optuna \citep{2019KDD....14.2623A}. Tuning uses a stratified 80\% of the
sources (3146 of 3933). Tuning used the nine-feature
set that included Galactic coordinates, and the seven-feature ensemble
reuses these hyperparameters without retuning.

\emph{Supervised classifiers.} Eight classifiers (decision tree, random
forest, SVC, AdaBoost, LightGBM, XGBoost, LR and LDA) are tuned over 100
trials with 3-fold stratified cross-validation. The Synthetic Minority Over-sampling Technique (SMOTE) \citep{Chawla_2002} is fitted on the training folds only. Each tuned
classifier is then refitted on 1000 random 80/20 splits of the pool, and
the three with the highest mean recall in each mode become the base
classifiers of CE, WE and BC.

\emph{PU classifiers.} Each PU classifier is tuned over 150 trials with
3-fold stratified cross-validation on the same pool, with no
oversampling. \citet{2024MNRAS.533.3283S} use one fixed
training-validation split; averaging over folds makes the tuned score
less dependent on a single split.

The objective is the Lee--Liu score \citep{Lee2003},
\begin{equation}
  L_2 = \frac{r^2}{\hat{p}_+},
  \label{eq:lee_liu}
\end{equation}
where $r$ is the recall on held-out known positives and $\hat{p}_+$ is
the fraction of all held-out sources predicted positive. It rewards
recall and penalises over-prediction, and it can be computed without
known negatives \citep{2024MNRAS.533.3283S}. It is bounded above by
$1/P(s=1)$, which is 4.80 in \fsrq{} mode and 2.64 in \bll{} mode, and
is only comparable between classifiers in the same mode on the same
data.

\subsection{Candidate Identification}
\label{sec:candidates_method}

\citet{2024MNRAS.533.3283S} retrain on a random 80\% of the data 1000
times and predict the whole catalogue. We score out-of-fold instead. In
each repeat, all sources are split into five stratified folds. Every
model is trained on four folds and scores the fifth, so that no source
is scored by a model trained on it. We repeat this 100 times with
different splits (200 times for the nine-feature run of
Section~\ref{sec:ablation_method}). In each repeat a \bcu is flagged if at least three of
the five classifiers label it positive. Its confidence is the fraction
of the repeats in which it was flagged. A \bcu is a candidate if its
confidence is at least 0.1, the same relative threshold as in
\citet{2024MNRAS.533.3283S} (more than 100 of 1000 iterations, so equivalent up to the boundary value). We also
report the counts at confidence $\geq 0.5$ and the effect of requiring 1
to 5 votes (Section~\ref{sec:pu_results}). The confidence measures how
stable the flag is across resampled splits. It is not a probability. We
assign each candidate to the highest tier it meets: A (confidence
$\geq 0.9$), B ($\geq 0.5$) or C ($\geq 0.1$). A \bcu is FSRQ-like or
BL~Lac-like if it is a candidate in one mode only, and ambiguous if it is
a candidate in both.

Held-out performance (recall and $L_2$) is computed from the same
out-of-fold predictions for the known sources.

\subsection{SHAP Feature Importance}
\label{sec:shap}

SHAP values \citep{Lundberg2017} give the contribution of each feature
to the score that a model assigns to a source. For each mode and each of
the five classifiers we fit the model once to all 3933 sources with the
tuned parameters and explain its positive-class score, not its hard
label. The score is the predicted probability where the classifier
provides one, and for PU Extra Trees the fraction of trees voting
positive. We use permutation SHAP on 600 sources, the same for every
model and stratified by class, with 30 background sources and 10
permutations per source. The importance of a feature is the mean
absolute SHAP value, averaged over the five models. Scores on different
scales weight the average unequally: BC in \fsrq{} mode scores only between
0.34 and 0.63, so it contributes little, and we also list the per-model
values.
\texttt{PL\_Index} and \texttt{log\_Pivot\_Energy} are strongly
correlated ($\rho=-0.85$), so each shares credit with the other and we
also report their combined contribution. SHAP describes what the fitted
models use. It does not show that a feature separates the classes physically. We also show a Uniform Manifold Approximation and Projection (UMAP) embedding of the standardised features, which displays the feature space, does not use the class labels and does not test the classification.

\subsection{Evaluation}
\label{sec:method_validation}

We evaluate the ensemble in five ways. Models fitted to the same
gamma-ray features cannot show that the \bcu~classes are correct, so
only the external checks (Sections~\ref{sec:external_method} to
\ref{sec:release_method}) bear on
that question.

\subsubsection{Labelled-only baselines}
\label{sec:baseline_method}

We train four two-class models on the known sources only (\fsrq~$=1$,
\bll~$=0$): logistic regression (LR), quadratic discriminant analysis
(QDA), a two-component Gaussian mixture per class (GMM2) and histogram
gradient boosting (HGB). They are evaluated with five repeats of
stratified five-fold cross-validation on the 2310 known sources, with
imputation and scaling fitted inside each training fold. We report the AUC, the recall and false-positive rate at a
score of 0.5, the recall at a false-positive rate of 0.08 (close to that
of the ensemble) and the Brier score. Each model is then fitted to all
known sources and scores the \bcus. These models inherit the selection
of the labelled sources, so agreement with the PU ensemble shows that
the features separate the classes, not that the \bcu~classes are
correct. As an unsupervised check, we fit Gaussian mixtures with one to
eight components (full covariance) to the known sources and compare the Bayesian information criterion (BIC) and the adjusted Rand index (ARI) against the true labels.

\subsubsection{Sensitivity to the sky coordinates}
\label{sec:ablation_method}

We refit the ensemble without Galactic longitude and latitude (GLON and GLAT), with the same folds, seeds
and tuned hyperparameters as the nine-feature run (the hyperparameters
are not retuned). The two runs are compared repeat by repeat over 100
repeats, using the first 100 repeats of the nine-feature run (so its
candidate counts differ slightly from those of the full 200-repeat run). The noise floor of the comparison is the overlap between two
independent halves of the nine-feature run (repeats 0 to 99 against 100
to 199), which differ in folds and seeds. A difference counts only if it
is clearly larger than this floor. We take the Galactic plane to be
$|b|<10^\circ$.

\subsubsection{Published optical spectra}
\label{sec:external_method}

We collect spectroscopic classifications of \fermi~sources from nine
papers: \citet{2016Ap&SS.361..316A}, \citet{2016AJ....151...95A},
\citet{PenaHerazo2017}, \citet{Marchesini2019},
\citet{2017MNRAS.467.2537K}, \citet{10.1093/mnras/stac2640},
\citet{AlvarezCrespo2025}, \citet{Marais2024} and
\citet{2019ApJS..241....5D}. A published class of FSRQ or BZQ (Roma-BZCAT class) is taken as FSRQ, and a class of BL~Lac, BZB or BZG (Roma-BZCAT classes) as BL~Lac. Sources classified
only as a quasar or a galaxy are dropped. The papers separate the
classes mostly with the rest-frame equivalent width limit of 5~\AA, and
a featureless spectrum is classified as BL~Lac. The BZQ label of \citet{AlvarezCrespo2025} is partly set by the radio spectrum and not by the emission-line width alone, so FSRQ labels from that paper are less strictly spectroscopic than the others. Sources are matched to
\fourFGL-DR4 by their \fourFGL, 3FGL (Third Source Catalog) or 3FHL (Third Catalog of Hard Sources) names (the latter two
through the association columns of the catalogue) or by position. If two
papers disagree on a source it is dropped. Two labelled entries are removed before matching: one 3FGL name that is
absent from the association columns of the catalogue, and
\fourFGL~J0746.5$-$0710, for which two papers give different classes. The
remaining labels cover 220 unique sources. Nine of these are not in the
score file, because they are not BCUs, FSRQs or BL~Lacs in \fourFGL-DR4 or
their names changed, which leaves 211 sources.

We split the matched sources into two groups. Group T1 are sources that
are still \bcus~in \fourFGL-DR4 and have a published class. This is the
independent test. Group T2 are sources with a published class and an
FSRQ or BL~Lac class in DR4. The DR4 class of these sources was often
derived from the same spectra, so T2 checks consistency and not accuracy.
At a confidence of 0.1 a source is \emph{correct} if it is a candidate in
the mode of its true class only, \emph{wrong} if it is a candidate in the
other mode only, \emph{ambiguous} if it is a candidate in both and
\emph{unclassified} if in neither. The precision is correct/(correct
$+$ wrong) and the coverage is (correct $+$ wrong)/$n$. We give Wilson
95\% intervals and the AUC of the difference between the FSRQ-mode and
BL~Lac-mode confidences.

\subsubsection{SDSS-V proxy classes}
\label{sec:sdssv_method}

The analysis of \citet{2026arXiv260726175I} gives pipeline spectral classes for \fermi~blazars from Sloan Digital Sky Survey V (SDSS-V) spectra
that are part of Data Release 20 \citep{SDSSDR20}.
The table is released as a DR20 value added catalogue
\citep{SDSSVAC2026}. We use the best-fit model class as a
proxy: sources fitted with emission lines or a quasi-stellar object (QSO) template are labelled
FSRQ and all others BL~Lac. Sources with redshift flags 1 or 2 and
sources with a spectral signal-to-noise ratio (S/N) below 5 are dropped.
On the known sources with $\mathrm{S/N}\geq5$ the proxy agrees with the
catalogue class for 107 of 127 \fsrqs~(84.3\%; 95\% interval 0.77 to
0.90) and 226 of 256 \blls~(88.3\%; 0.84 to 0.92), so the proxy has an
error of 12 to 16\%. If a source also has a published class, the
published class is used.

\subsubsection{\bcus~classified in later releases}
\label{sec:release_method}

A source that was a \bcu~in an earlier release and has an FSRQ or
BL~Lac class in \fourFGL-DR4 provides a test in which the later class is
the answer. We use \fourFGL~Data Release 1 (DR1) \citep{2020ApJS..247...33A} and Data Release 2 (DR2) \citep{2020arXiv200511208B}, because Data Release 3 (DR3) \citep{2022ApJS..260...53A}
supplies only 7 such sources. Names are matched between releases and
unmatched sources are left out. The scores are the out-of-fold ones, so
each source is scored by models that never saw it. The hyperparameters,
however, were tuned on a random 80\% of all sources, which includes
most of these. The DR2
test set largely overlaps the DR1 one, so we treat DR2 as a consistency
check on DR1.

\section{Results}
\label{sec:results}

All results below are for the seven-feature ensemble (no Galactic
coordinates) unless stated otherwise. Confidences are the fraction of
100 out-of-fold repeats in which at least three of five classifiers flag
a source. They measure how stable a flag is across training splits and
are not probabilities that the assigned class is correct.

\subsection{PU Ensemble Performance}
\label{sec:pu_results}

\begin{figure}[ht]
\centering
\includegraphics[width=\columnwidth]{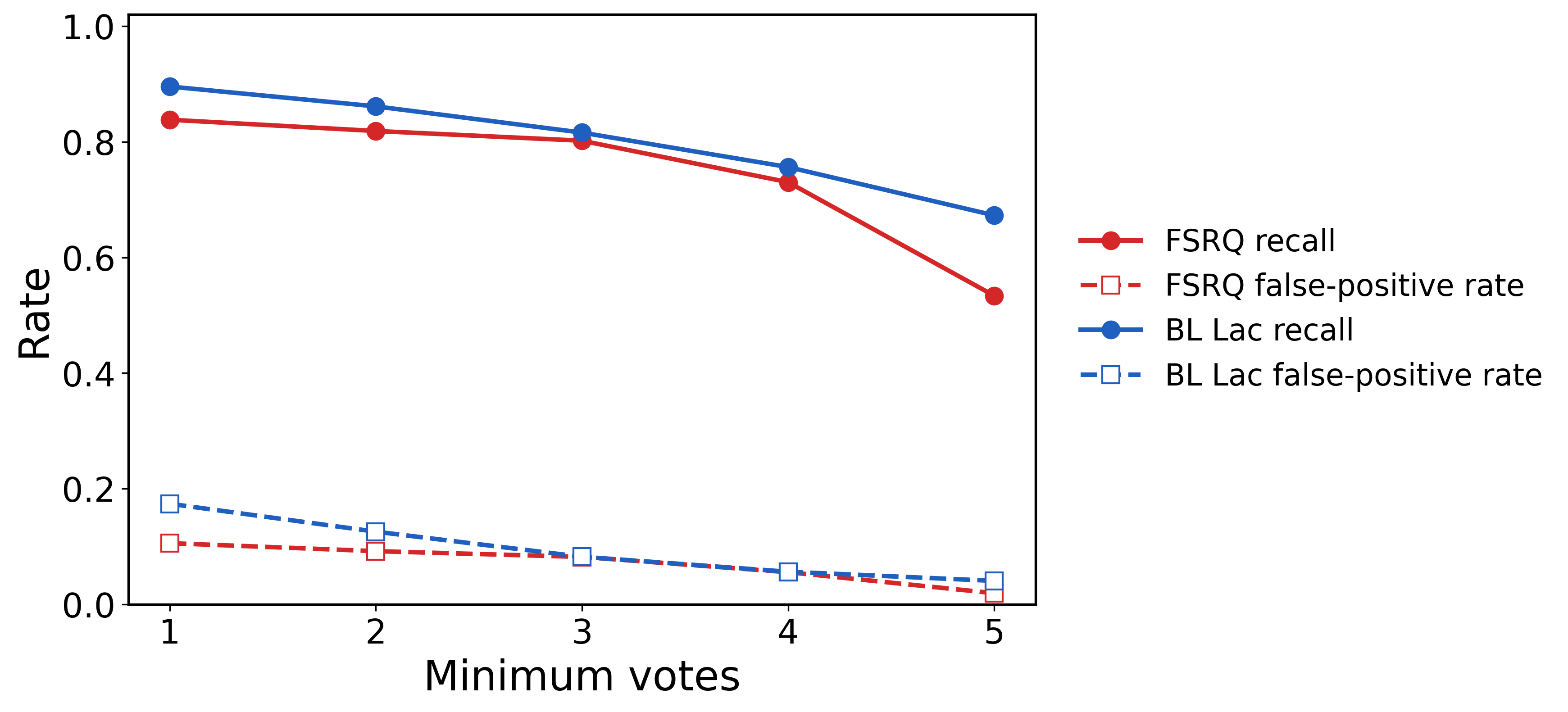}
\caption{
  Held-out recall (filled circles) and false-positive rate (open squares) against the minimum number of votes, for \fsrq{} mode (red) and \bll{} mode (blue). Values are means over 100 out-of-fold repeats on the known sources.
}
\label{fig:vote_sweep}
\end{figure}

Table~\ref{tab:pu_results} gives the held-out performance of each
classifier and of the vote rule, computed from the out-of-fold
predictions for the known sources and averaged over the repeats. The
vote rule (at least three of five classifiers) has a recall of
$0.802\pm0.005$ at a false-positive rate of $0.082\pm0.002$ in \fsrq{}
mode, and a recall of $0.816\pm0.004$ at $0.082\pm0.003$ in \bll{} mode.
The false-positive rate is the share of known sources of the other class
that are flagged. Among the single classifiers, MLR is the most
conservative, with the lowest recall (0.54 and 0.69) and the lowest
false-positive rate (0.023 and 0.048), while CE and WE flag the most
sources in both modes, and BC does so in \fsrq{} mode. The vote rule has
the highest $L_2$ score in \fsrq{} mode (2.14) and ties with CE in \bll{}
mode (1.43), against upper bounds of 4.80 and 2.64.
Table~\ref{tab:vote} and Figure~\ref{fig:vote_sweep} show how the
threshold on the number of votes trades recall against false positives.
Requiring four votes lowers the false-positive rate to 0.056 and 0.057
and the recall to 0.73 and 0.76. We keep three votes, the rule of
\citet{2024MNRAS.533.3283S}.

\begin{deluxetable*}{lcccccc}
\tablecaption{
  Held-out performance of the five PU classifiers and of the vote rule on the known sources, in \fsrq{} mode (labelled positives are the \fsrqs) and in \bll{} mode (labelled positives are the \blls), averaged over 100 out-of-fold repeats (mean $\pm$ standard deviation over repeats). Recall is the share of the labelled positives that are flagged, FPR the false-positive rate, and $L_2$ the Lee--Liu score.
  \label{tab:pu_results}
}
\tablewidth{0pt}
\tablehead{
  \colhead{} &
  \multicolumn{3}{c}{\fsrq{} mode} &
  \multicolumn{3}{c}{\bll{} mode} \\
  \colhead{Classifier} &
  \colhead{Recall} & \colhead{FPR} & \colhead{$L_2$} &
  \colhead{Recall} & \colhead{FPR} & \colhead{$L_2$}
}
\startdata
Classic Elkanoto  & $0.811\pm0.006$ & $0.087\pm0.003$ & 2.122
                  & $0.853\pm0.003$ & $0.136\pm0.004$ & 1.431 \\
Weighted Elkanoto & $0.803\pm0.005$ & $0.091\pm0.002$ & 2.107
                  & $0.883\pm0.003$ & $0.134\pm0.004$ & 1.419 \\
Bagging PU        & $0.822\pm0.007$ & $0.096\pm0.002$ & 2.104
                  & $0.777\pm0.002$ & $0.058\pm0.002$ & 1.415 \\
MLR               & $0.543\pm0.024$ & $0.023\pm0.003$ & 1.829
                  & $0.695\pm0.023$ & $0.048\pm0.006$ & 1.371 \\
PU Extra Trees    & $0.744\pm0.011$ & $0.060\pm0.003$ & 2.071
                  & $0.793\pm0.011$ & $0.104\pm0.007$ & 1.366 \\
Vote $\geq3$      & $0.802\pm0.005$ & $0.082\pm0.002$ & 2.141
                  & $0.816\pm0.004$ & $0.082\pm0.003$ & 1.431 \\
\enddata
\tablecomments{
  FPR: share of the known sources of the other class that are flagged. $L_2=r^2/\hat{p}_+$ is the Lee--Liu score, where $r$ is the recall and $\hat{p}_+$ the fraction of all held-out sources predicted positive; it is bounded above by 4.80 in \fsrq{} mode and 2.64 in \bll{} mode. Vote $\geq3$: a source is flagged if at least three of the five classifiers flag it.
}
\end{deluxetable*}

\begin{deluxetable}{llccc}
\tablecaption{
  Effect of the vote threshold on the held-out performance on the known sources, in \fsrq{} mode and in \bll{} mode. Votes is the number of the five classifiers that must flag a source (5 means all five), Recall the share of the labelled positives flagged, FPR the share of the known sources of the other class flagged, and BCUs flagged the number of \bcus~flagged.
  \label{tab:vote}
}
\tablewidth{0pt}
\tablehead{
  \colhead{Mode} & \colhead{Votes} & \colhead{Recall} &
  \colhead{FPR} & \colhead{BCUs flagged}
}
\startdata
\fsrq{} & $\geq1$ & 0.838 & 0.106 & 487 \\
        & $\geq2$ & 0.819 & 0.092 & 437 \\
        & $\geq3$ & 0.802 & 0.082 & 401 \\
        & $\geq4$ & 0.730 & 0.056 & 316 \\
        & $5$     & 0.534 & 0.019 & 147 \\
\hline
\bll{}  & $\geq1$ & 0.895 & 0.174 & 774 \\
        & $\geq2$ & 0.861 & 0.126 & 658 \\
        & $\geq3$ & 0.816 & 0.082 & 546 \\
        & $\geq4$ & 0.756 & 0.057 & 422 \\
        & $5$     & 0.673 & 0.041 & 280 \\
\enddata
\tablecomments{
  BCUs flagged is the mean number of \bcus~flagged in one repeat, not the
  number of candidates.
}
\end{deluxetable}

\subsection{Labelled-Only Baselines}
\label{sec:baseline}

Table~\ref{tab:baseline} compares the ensemble with classifiers trained
on the known \fsrqs~and \blls~only. Logistic regression separates the two
classes best, with an AUC of $0.953$, a recall of $0.850$ at a
false-positive rate of $0.077$ and a recall of $0.853$ at a rate of
0.08. The ensemble in \fsrq{} mode reaches $0.802$ at $0.082$. On known
sources the PU ensemble therefore does not beat a simple two-class model
trained on the same labels. Both methods use the same selected set of
labelled sources, so this comparison says nothing about their accuracy on
\bcus.

The logistic regression agrees with the ensemble on the \bcus~that the
ensemble classifies. It gives $P\geq0.5$ to 98\% of the FSRQ-like
candidates and $P<0.5$ to all of the BL~Lac-like candidates, and the
rank correlation of its score with the ensemble's confidence difference
is 0.90. It gives an intermediate score ($0.1<P<0.9$) to 87\% of the
\bcus~that the ensemble leaves unclassified, against 50\% of all \bcus.
The QDA, GMM2 and HGB models agree less well. They give $P\geq0.5$ to
96\%, 87\% and 96\% of the FSRQ-like candidates and an intermediate score
to 57\%, 32\% and 78\% of the unclassified \bcus, and their rank
correlations with the ensemble are 0.85, 0.81 and 0.89.
The ensemble's abstention therefore falls mostly where the two-class
model is also uncertain.

An unsupervised Gaussian mixture does not recover the two classes. The
BIC keeps falling up to seven components (13\,725 at seven, 34\,011 at
one), and the best ARI against the labels is 0.13. With two components,
one holds 613 of the 820 \fsrqs~but also 575 of the 1490 \blls. The
two classes overlap in these features, and the data do not show two
well-separated clusters.

\begin{deluxetable*}{lccccc}
\tablecaption{
  Labelled-only classifiers, evaluated with five repeats of five-fold
  cross-validation on the 2310 known sources (\fsrq{} is the positive class). AUC is the area under the ROC curve. Recall at 0.5 and FPR at 0.5 are the share of \fsrqs~and of \blls~called \fsrq{} when the score threshold is 0.5. Recall at FPR$=0.08$ is the recall at the threshold that gives a false-positive rate of 0.08. Brier is the mean squared difference between the predicted probability and the true label (lower is better). The last row gives the PU ensemble in \fsrq{} mode for comparison; its recall and false-positive rate are those of the vote rule, not of a score threshold of 0.5. \label{tab:baseline}
}
\tablewidth{0pt}
\tablehead{
  \colhead{Model} & \colhead{AUC} & \colhead{Recall at 0.5} &
  \colhead{FPR at 0.5} & \colhead{Recall at FPR$=0.08$} &
  \colhead{Brier}
}
\startdata
LR   & 0.953 & $0.850\pm0.001$ & $0.077\pm0.001$ & $0.853\pm0.002$ & 0.077 \\
QDA  & 0.936 & $0.828\pm0.001$ & $0.086\pm0.001$ & $0.818\pm0.004$ & 0.091 \\
GMM2 & 0.937 & $0.785\pm0.003$ & $0.064\pm0.002$ & $0.825\pm0.004$ & 0.094 \\
HGB  & 0.946 & $0.833\pm0.005$ & $0.074\pm0.003$ & $0.838\pm0.006$ & 0.083 \\
\hline
PU ensemble (\fsrq{} mode, vote $\geq3$) & --- & $0.802$ & $0.082$ & --- & --- \\
\enddata
\tablecomments{
  The standard deviation of the AUC over repeats is below 0.001 for all
  models. LR: logistic regression; QDA: quadratic discriminant analysis;
  GMM2: two-component Gaussian mixture per class; HGB: histogram gradient
  boosting.
}
\end{deluxetable*}

\subsection{BCU Candidates}
\label{sec:cand_results}

\begin{figure}[ht]
\centering
\includegraphics[width=\columnwidth]{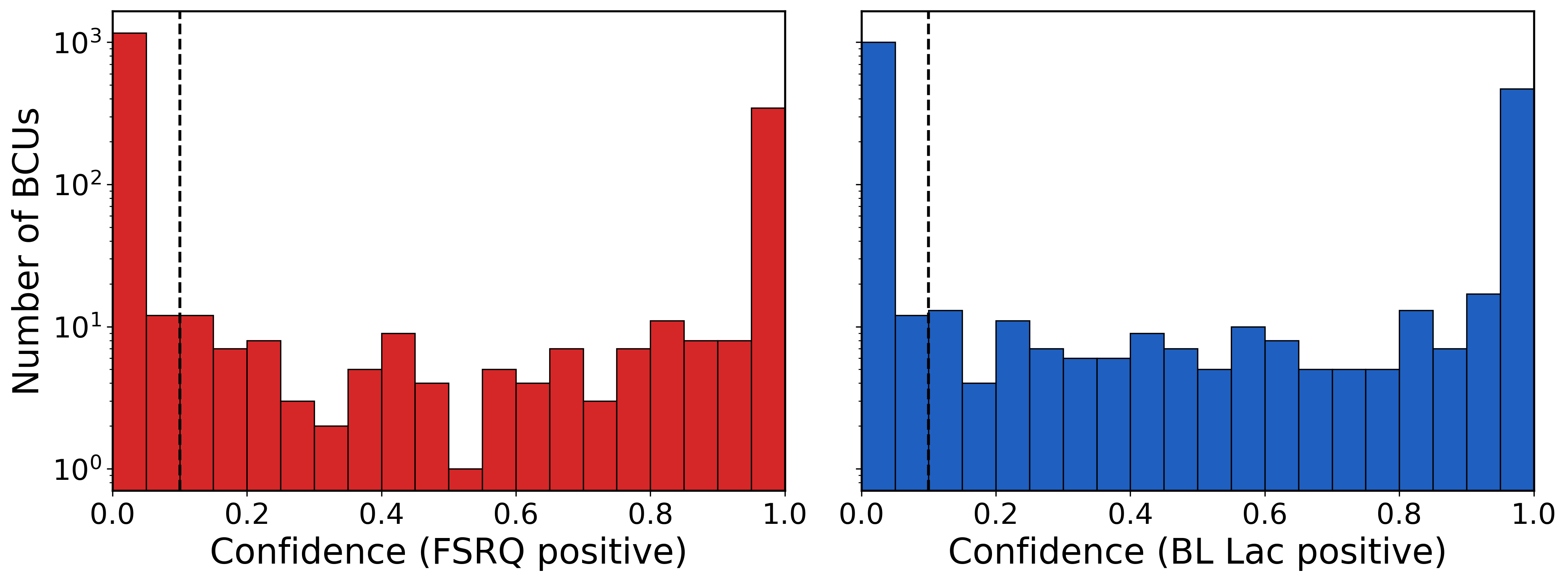}
\caption{
  Distribution of the confidence of the \bcus~in \fsrq{} mode (left) and
  \bll{} mode (right), on a logarithmic count axis. Confidence is the fraction of the 100 out-of-fold repeats in which at least three of the five classifiers flag a source. The dashed line is the candidate threshold of 0.1.
}
\label{fig:confidence}
\end{figure}

At a confidence of at least 0.1, 1046 of the 1623 \bcus~(64\%) are
candidates in at least one mode (Table~\ref{tab:candidates}).
A \bcu~that is a candidate in one mode only is assigned to that class: 437
are FSRQ-like and 597 are BL~Lac-like. Twelve sources are candidates in
both modes and are called ambiguous, and 577 are candidates in neither.
Of the FSRQ-like candidates, 345 (79\%) have a confidence of at least 0.9
(tier A), 43 are in tier B (0.5 to 0.9) and 49 in tier C (0.1 to 0.5).
For the BL~Lac-like candidates the numbers are 484 (81\%), 54 and 59
(Table~\ref{tab:tiers}). Figure~\ref{fig:confidence} shows the
distribution of the confidence.

\begin{deluxetable*}{lrrrrrr}
\tablecaption{
  \bcu~candidates at three confidence thresholds. Confidence is the threshold applied to the confidence of each mode. \fsrq{} mode and \bll{} mode are the numbers of \bcus~at or above the threshold in that mode; \fsrq{} only and \bll{} only are the candidates in one mode only; Both are the candidates in both modes; Neither are the \bcus~below the threshold in both modes. \label{tab:candidates}
}
\tablewidth{0pt}
\tablehead{
  \colhead{Confidence} &
  \colhead{\fsrq{} mode} & \colhead{\bll{} mode} &
  \colhead{\fsrq{} only} & \colhead{\bll{} only} &
  \colhead{Both} & \colhead{Neither}
}
\startdata
$\geq0.1$ & 449 & 609 & 437 & 597 & 12 & 577 \\
$\geq0.5$ & 399 & 546 & 392 & 539 &  7 & 685 \\
$\geq0.9$ & 353 & 488 & 350 & 485 &  3 & 785 \\
\enddata
\tablecomments{
  Out of 1623 \bcus. ``\fsrq{} mode'' counts all \bcus~above the
  threshold in that mode, including those that are also candidates in the
  other mode.
}
\end{deluxetable*}

\begin{deluxetable}{lrrrr}
\tablecaption{
  Number of candidates in each stability tier. Class is the class assigned at a confidence of 0.1; A, B and C are the tiers of the confidence in that class; Total is the number of candidates (ambiguous sources are candidates in both modes and have no tier, and no-class \bcus~are below 0.1 in both modes). \label{tab:tiers}
}
\tablewidth{0pt}
\tablehead{
  \colhead{Class} & \colhead{A} & \colhead{B} & \colhead{C} &
  \colhead{Total}
}
\startdata
FSRQ-like   & 345 & 43 & 49 & 437 \\
BL~Lac-like & 484 & 54 & 59 & 597 \\
Ambiguous   & \multicolumn{3}{c}{} & 12 \\
No class    & \multicolumn{3}{c}{} & 577 \\
\enddata
\tablecomments{
  Tier A: confidence $\geq0.9$; B: $0.5$ to $0.9$; C: $0.1$ to $0.5$.
}
\end{deluxetable}

The candidates differ from the labelled sources in the same ways as the
\bcus~as a whole (Table~\ref{tab:selection}). The \bcus~without a class
are the faintest group (median $\log F_{1000}=-9.77$, against $-9.40$
and $-9.51$ for the known \fsrqs~and \blls) and the one closest to the
plane (25\% at $|b|<10^\circ$, against 4.5\% and 5.6\%). The FSRQ-like
candidates have a softer median photon index than the known \fsrqs~(2.60
against 2.47) and are fainter by 0.24~dex. The candidates are selected
with these features, so these differences are in part a consequence of
the selection and do not test the classes.

\begin{deluxetable*}{lrcccc}
\tablecaption{
  Properties of the labelled sources and of the \bcu~groups. Group is the known \fsrqs, the known \blls, or the \bcus~by assigned class at a confidence of 0.1; $n$ is the number of sources; Share at $|b|<10^\circ$ is the fraction of the group at Galactic latitude $|b|<10^\circ$; the last three columns are the group medians of the integral photon flux, the pivot energy and the photon index. \label{tab:selection}
}
\tablewidth{0pt}
\tablehead{
  \colhead{Group} & \colhead{$n$} & \colhead{Share at $|b|<10^\circ$} &
  \colhead{Median $\log F_{1000}$} &
  \colhead{Median $\log E_\mathrm{pivot}$} &
  \colhead{Median \texttt{PL\_Index}}
}
\startdata
Known \fsrqs     &  820 & 0.045 & $-9.404$ & 2.915 & 2.472 \\
Known \blls      & 1490 & 0.056 & $-9.510$ & 3.399 & 2.018 \\
\bcus, FSRQ-like &  437 & 0.146 & $-9.641$ & 2.925 & 2.596 \\
\bcus, BL~Lac-like & 597 & 0.151 & $-9.746$ & 3.551 & 2.023 \\
\bcus, no class  &  577 & 0.253 & $-9.772$ & 3.247 & 2.348 \\
All \bcus        & 1623 & 0.185 & $-9.735$ & 3.256 & 2.295 \\
\enddata
\tablecomments{
  The twelve ambiguous \bcus~are included in ``All \bcus'' only.
  $F_{1000}$ is the integral photon flux in ph\,cm$^{-2}$\,s$^{-1}$ and
  $E_\mathrm{pivot}$ the pivot energy in MeV.
}
\end{deluxetable*}

An earlier run of the same method (nine features, 1000 retrains of 80\%
of the data) listed 880 candidates. Of these, 859 (98\%) are candidates
here, 846 with the same class, 3 with the opposite class and 10 as
ambiguous, and 82\% of the present candidates were in the earlier list. This shows that the
lists are stable against changes in the training scheme. It is not
independent confirmation, because both runs use the same data, and
the seven features used here are a subset of the nine used earlier.

\subsection{Dependence on the Sky Coordinates}
\label{sec:ablation}

We compare the seven-feature ensemble with the nine-feature one
(Section~\ref{sec:ablation_method}). Table~\ref{tab:ablation} gives the
held-out performance of both. Removing the coordinates changes the recall of
the vote rule by $+0.013$ in \fsrq{} mode and $-0.007$ in \bll{} mode, and
the false-positive rate by $+0.004$ and $-0.005$, so the coordinates do
not improve the performance on the known sources. In \fsrq{} mode, WE is
the only classifier whose performance changes appreciably: its recall rises
from 0.772 to 0.803 and its false-positive rate from 0.077 to 0.091, while
the vote recall rises by 0.013. In \bll{} mode, BC, MLR and PUET change by
0.013 to 0.018 in recall or false-positive rate, while the vote rule
changes by less than 0.01.

\begin{deluxetable*}{llcccccc}
\tablecaption{
  Held-out performance of each classifier (Classic Elkanoto CE, Weighted Elkanoto WE, Bagging PU BC, Modified Logistic Regression MLR, PU Extra Trees PUET) and of the vote rule, with and without the Galactic coordinates (nine and seven features), on the same 100 repeats, in \fsrq{} mode and in \bll{} mode. Recall is the share of the labelled positives flagged and FPR the share of the known sources of the other class flagged. \label{tab:ablation}
}
\tablewidth{0pt}
\tablehead{
  \colhead{Mode} & \colhead{Classifier} &
  \multicolumn{2}{c}{Recall} & \multicolumn{2}{c}{FPR} &
  \multicolumn{2}{c}{BCUs flagged} \\
  \colhead{} & \colhead{} &
  \colhead{9} & \colhead{7} & \colhead{9} & \colhead{7} &
  \colhead{9} & \colhead{7}
}
\startdata
\fsrq{} & CE        & 0.812 & 0.811 & 0.088 & 0.087 & 410 & 425 \\
        & WE        & 0.772 & 0.803 & 0.077 & 0.091 & 354 & 411 \\
        & BC        & 0.826 & 0.822 & 0.100 & 0.096 & 429 & 446 \\
        & MLR       & 0.543 & 0.543 & 0.023 & 0.023 & 157 & 154 \\
        & PUET      & 0.752 & 0.744 & 0.060 & 0.060 & 336 & 351 \\
        & Vote $\geq3$ & 0.789 & 0.802 & 0.078 & 0.082 & 369 & 401 \\
\hline
\bll{}  & CE        & 0.857 & 0.853 & 0.138 & 0.136 & 580 & 617 \\
        & WE        & 0.884 & 0.883 & 0.139 & 0.134 & 700 & 735 \\
        & BC        & 0.791 & 0.777 & 0.071 & 0.058 & 440 & 473 \\
        & MLR       & 0.709 & 0.695 & 0.061 & 0.048 & 293 & 311 \\
        & PUET      & 0.811 & 0.793 & 0.096 & 0.104 & 471 & 545 \\
        & Vote $\geq3$ & 0.823 & 0.816 & 0.087 & 0.082 & 497 & 546 \\
\enddata
\tablecomments{
  Columns headed 9 and 7 give the values with nine and seven features.
  BCUs flagged is the mean number of \bcus~flagged in one repeat.
}
\end{deluxetable*}

The \bcu~scores change much more than the known-source performance
(Table~\ref{tab:overlap}). In \fsrq{} mode the seven-feature run keeps
98.8\% of the nine-feature candidates and adds 6.9\% new ones, against
98.3\% and 0.7\% for two independent halves of the nine-feature run. In
\bll{} mode it keeps 94.1\% and adds 13.6\% new, against 98.4\% and 0.5\%.
The rank correlation of the confidences is 0.96 and 0.92, against 0.99 for
the noise floor. At a confidence of 0.1 the share of new candidates is about 10 and 25
times the noise floor. No \bcu~changes from FSRQ-like to BL~Lac-like or the reverse
(Table~\ref{tab:assign}). In total, 1472 \bcus~keep their status, 107 gain a
class, 30 lose one and 14 move between a single class and the ambiguous
status.
Most of the gained candidates are weak: 56\% are in tier C and 17\% in
tier A. Among the 920 \bcus~with the same class in both runs, 119 move to a
higher tier and 30 to a lower one, so the seven-feature run raises the
confidence of many existing candidates as well.

\begin{deluxetable*}{llcccccc}
\tablecaption{
  Overlap of the candidate sets (confidence $\geq$ the threshold) between
  the nine- and seven-feature runs, and between two independent halves of
  the nine-feature run (noise floor). The nine-feature candidates are those
  of its first 100 repeats, the same number as in the seven-feature run.
  \label{tab:overlap}
}
\tablewidth{0pt}
\tablehead{
  \colhead{Mode} & \colhead{Threshold} &
  \colhead{$n$ (9)} & \colhead{$n$ (7)} &
  \colhead{Kept} & \colhead{New} &
  \colhead{Kept (floor)} & \colhead{New (floor)}
}
\startdata
\fsrq{} & 0.1 & 423 & 449 & 0.988 & 0.069 & 0.983 & 0.007 \\
        & 0.5 & 372 & 399 & 0.984 & 0.083 & 0.987 & 0.011 \\
        & 0.9 & 319 & 353 & 0.991 & 0.105 & 0.987 & 0.022 \\
\hline
\bll{}  & 0.1 & 559 & 609 & 0.941 & 0.136 & 0.984 & 0.005 \\
        & 0.5 & 489 & 546 & 0.949 & 0.150 & 0.992 & 0.006 \\
        & 0.9 & 443 & 488 & 0.950 & 0.137 & 0.991 & 0.007 \\
\enddata
\tablecomments{
  Kept: share of the nine-feature candidates also found with seven
  features. New: share of the seven-feature candidates that are not in the
  nine-feature set. The floor columns compare repeats 0 to 99 with repeats 100 to 199 of the nine-feature run. $n$ (9) and $n$ (7) are the numbers of candidates in the nine- and seven-feature runs, and Threshold is the confidence threshold.
}
\end{deluxetable*}

\begin{deluxetable}{lrrrr}
\tablecaption{
  Class assignment of the 1623 \bcus~with nine features (rows) and seven features (columns). Each entry is the number of \bcus~with that pair of assignments; Both means a candidate in both modes (ambiguous) and No class a candidate in neither mode. \label{tab:assign}
}
\tablewidth{0pt}
\tablehead{
  \colhead{} & \colhead{BL~Lac-like} & \colhead{FSRQ-like} &
  \colhead{Both} & \colhead{No class}
}
\startdata
BL~Lac-like & 520 &   0 &  0 &  26 \\
FSRQ-like   &   0 & 400 &  7 &   3 \\
Both        &   1 &   6 &  5 &   1 \\
No class    &  76 &  31 &  0 & 547 \\
\enddata
\end{deluxetable}

The shift is largest for the \bcus~in the Galactic plane. With nine
features, 55.0\% of the 300 \bcus~at $|b|<10^\circ$ receive no class,
against 37.0\% at $|b|\geq10^\circ$, a difference of 18.0 percentage
points. With seven features the numbers are 48.7\% and 32.6\%, a
difference of 16.1 points (Table~\ref{tab:plane}). Removing the
coordinates therefore explains about 2 of the 18 points. The remaining
excess of unclassified \bcus~in the plane is not explained by the
coordinates. It could come from the lower flux of these sources, from a
different population in the plane, or from both, and we cannot separate
these with the present data.

\begin{deluxetable}{llcccc}
\tablecaption{
  Class assignment of the \bcus~in and out of the Galactic plane. Features is the number of features of the run, Region the range of Galactic latitude $|b|$, $n$ the number of \bcus~in the region, and the last three columns the fraction of those \bcus~assigned to each class. \label{tab:plane}
}
\tablewidth{0pt}
\tablehead{
  \colhead{Features} & \colhead{Region} & \colhead{$n$} &
  \colhead{FSRQ-like} & \colhead{BL~Lac-like} & \colhead{No class}
}
\startdata
Nine  & $|b|<10^\circ$      &  300 & 0.177 & 0.273 & 0.550 \\
      & $|b|\geq10^\circ$   & 1323 & 0.270 & 0.351 & 0.370 \\
Seven & $|b|<10^\circ$      &  300 & 0.213 & 0.300 & 0.487 \\
      & $|b|\geq10^\circ$   & 1323 & 0.282 & 0.383 & 0.326 \\
\enddata
\tablecomments{
  Fractions of the \bcus~in each region. The small fraction of ambiguous
  sources (0 to 0.010) is not listed.
}
\end{deluxetable}

\subsection{External Checks}
\label{sec:external}

The ensemble is trained only on the known \fsrqs~and \blls, so its
assignments for the \bcus~can be tested only against classes that come from
elsewhere. We use four external sets (Sections~\ref{sec:external_method} to
\ref{sec:release_method}): published optical spectra, SDSS-V spectra, and the
classes that \bcus~of the earlier releases DR1 and DR2 received in DR4. In each set we take the \bcus~that
the ensemble assigns to a single class at a confidence of 0.1 and ask how
often that class agrees with the external one. We report this precision for
FSRQ-like and BL~Lac-like calls separately, the coverage (the share of the
sample that receives a single-class call), and the AUC of the difference
between the \fsrq-mode and \bll-mode confidences, which does not depend on
the 0.1 threshold. Table~\ref{tab:external} gives the summary. The samples
come from spectroscopic follow-up or from later reclassification, so they are
selected and do not represent all \bcus; the points on selection are collected
at the end of this section.

\subsubsection{Published optical spectra}

Forty-two \bcus~that are still \bcus~in DR4 have a published class: 19
\fsrqs~and 23 \blls~(group T1; Section~\ref{sec:external_method}). This is the
cleanest test, because the classes come from spectra that did not enter the
training and are not part of the DR4 classification. Of the 28 sources that
the ensemble assigns to one class, 25 are correct (89\%; 95\% interval 0.73
to 0.96) and 3 are wrong. All three errors are \blls~called FSRQ-like. The
FSRQ-like calls are correct for 15 of 18 (83\%; 0.61 to 0.94) and the
BL~Lac-like calls for 10 of 10 (100\%; 0.72 to 1.00). The coverage differs
strongly between the classes. The ensemble assigns a class to 15 of the 19
\fsrqs~(all correct) and to 13 of the 23 \blls~(10 correct), so 14 of the 42
sources get no class, 10 of them \blls. No source is flagged in both modes.
The AUC of the confidence difference is 0.89.

The 169 sources in group T2 have a published class and a DR4 class.
The two agree for 166 of them. The three disagreements are two \blls~in DR4
whose spectra give an FSRQ and one DR4 \fsrq~whose spectrum gives a
BL~Lac. The ensemble, with the scores out of fold from the seven-feature
classifiers, assigns the published class for 17 of the 23 \fsrqs~(74\%; 0.54
to 0.88) and 131 of the 146 \blls~(90\%; 0.84 to 0.94), with an AUC of 0.93.
It assigns the other class to 3 \fsrqs~and 6 \blls, and no class to 3 and 9.
This is a consistency check, because the DR4 classes of these sources rest
partly on the same spectra. It shows that the ensemble reproduces the
catalogue classes of sources it has not seen when they are scored out of fold,
but it does not test the \bcus.

\subsubsection{SDSS-V proxy classes}

For 89 \bcus~with SDSS-V spectra (S/N $\geq5$; 31 proxy \fsrqs~and 58
proxy \blls) the ensemble assigns a single class to 70, of which 57 agree
with the proxy (81\%; 0.71 to 0.89). The FSRQ-like calls are correct for 20
of 26 (77\%; 0.58 to 0.89) and the BL~Lac-like calls for 37 of 44 (84\%; 0.71
to 0.92). The AUC is 0.79, lower than for the published spectra. The proxy
itself is not exact. It disagrees with the catalogue class for 12 to 16\% of
the known sources (Section~\ref{sec:sdssv_method}), so some of the 13
disagreements may be errors of the proxy and not of the ensemble. Four of the
31 proxy \fsrqs~and 15 of the 58 proxy \blls~have no class. As for the
published spectra, the unclassified sources are mostly \blls, but the share is
smaller for the \fsrqs~than for the \blls~in both samples.

\subsubsection{\bcus~classified in later releases}

\begin{figure*}[ht]
\centering
\includegraphics[width=0.48\textwidth]{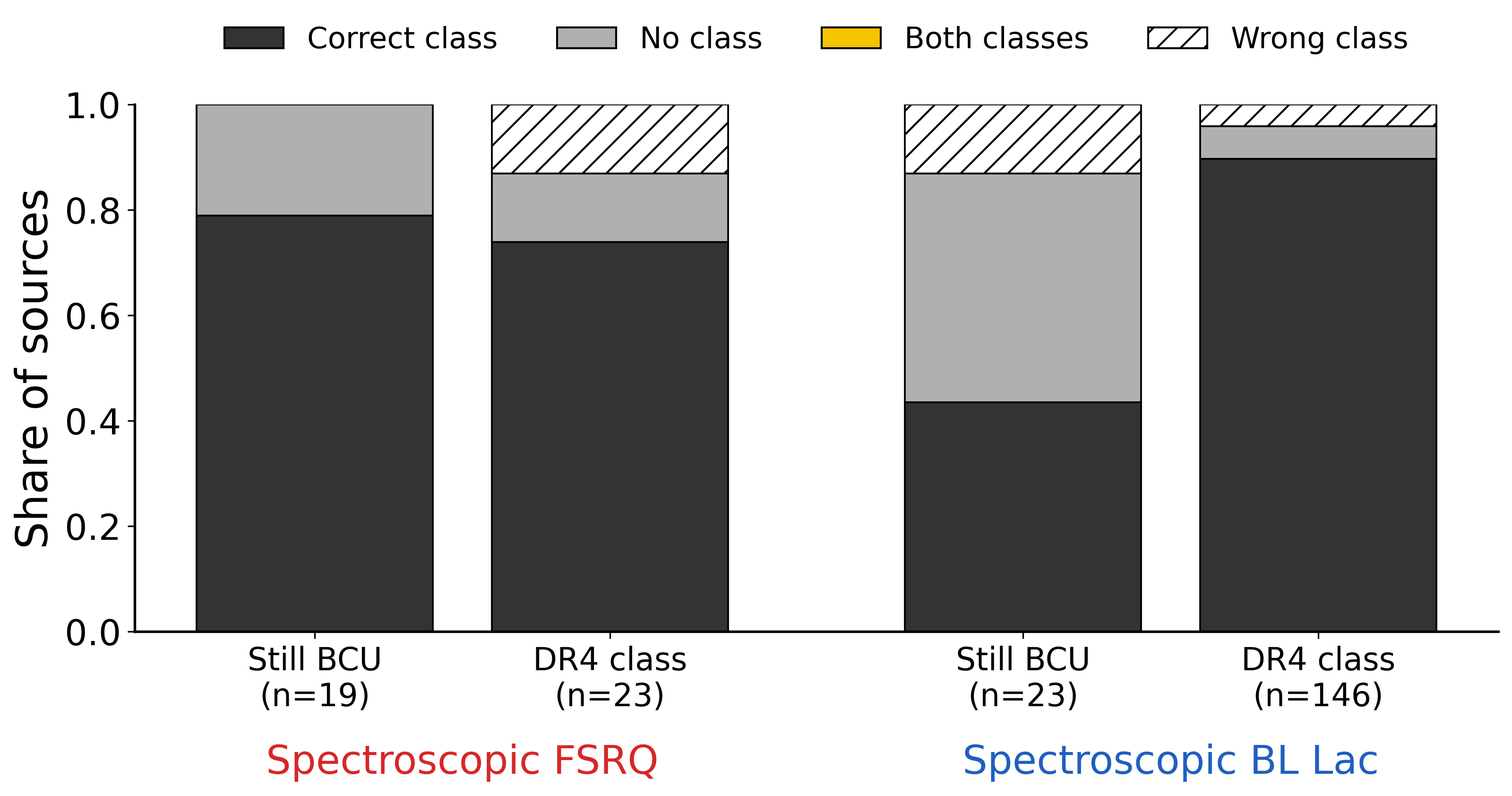}
\includegraphics[width=0.48\textwidth]{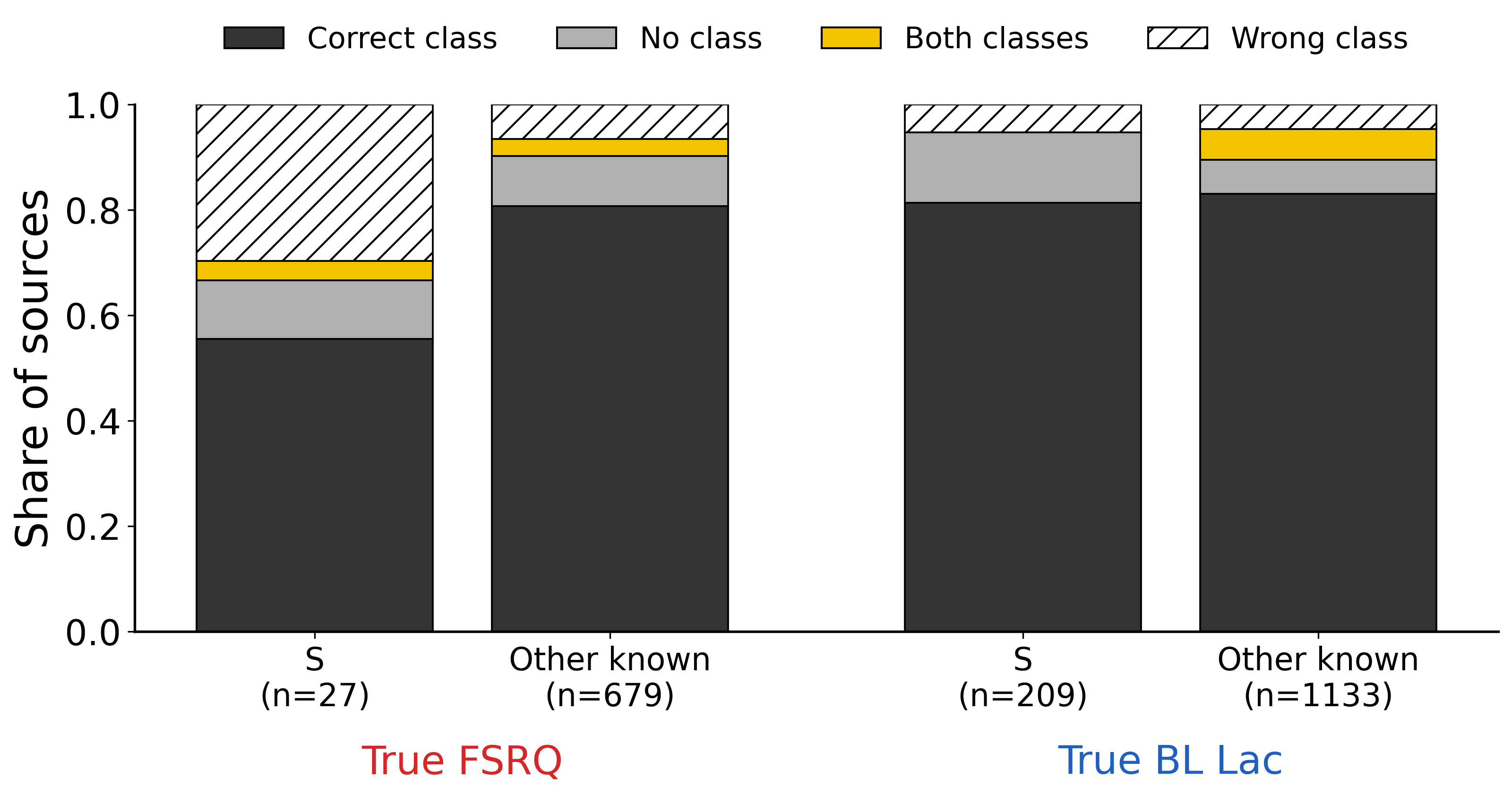}
\caption{
  Share of sources assigned to the correct class, to no class, to both
  classes and to the wrong class, for true \fsrqs~and \blls. \textit{Left}:
  \bcus~with a published class. \textit{Right}: \bcus~in DR1 that were classified in DR4 (S), and the other known sources of DR1 for comparison. On the left, Still BCU is group T1 and DR4 class is group T2 (Section~\ref{sec:external_method}).
}
\label{fig:dr1}
\end{figure*}

Figure~\ref{fig:dr1} and Table~\ref{tab:external} give the result for
\bcus~in DR1 or DR2 that are \fsrqs~or \blls~in DR4. These sources form the
largest of the four samples. For DR1 (236 sources: 27 \fsrqs, 209
\blls; 80\% of the DR4 sources are matched) the ensemble assigns a single
class to 204. The BL~Lac-like calls are correct for 170 of 178 (96\%; 0.91
to 0.98), and the FSRQ-like calls for 15 of 26 (58\%; 0.39 to 0.74). The two classes behave differently when we ask how many sources of each true class the ensemble finds. For the 209 \blls, the BL~Lac-like call is made
for 81\%, close to the 83\% for the known \blls. For the 27 \fsrqs, the
FSRQ-like call is made for only 15 (56\%), against 81\% for the known
\fsrqs; 8 are called BL~Lac-like and 4 receive no single class. The AUC is 0.83,
against 0.94 for the known sources. DR2 (114 sources: 15 \fsrqs, 99 \blls; 87\%
matched) gives the same picture, with 8 of 15 \fsrqs~found and 4 called
BL~Lac-like; its BL~Lac-like calls are correct for 80 of 84 (95\%), and its
AUC is again 0.83, against 0.94 for the other known sources. The DR2 set is
largely a subset of the DR1 set, so the two are not independent confirmations.

The FSRQ-like precision is low in part because \blls~outnumber \fsrqs~by
about 8 to 1 in the DR1 group, so a small false-positive rate on \blls~gives many
wrong FSRQ-like calls: 11 of the 209 \blls~(5\%) are called FSRQ-like, which
is enough to supply 11 of the 26 FSRQ-like calls. That ratio reflects which
\bcus~were classified, not the \bcu~population.

\subsubsection{Summary of the external checks}

In all four samples the BL~Lac-like calls are the more reliable ones (84 to
100\% correct, against 58 to 83\% for the FSRQ-like calls), and in all four
the intervals of the FSRQ-like precision are wide because only 12 to 26
sources are involved. The ensemble leaves 14 to 33\% of each sample
unclassified (coverage 67 to 86\%), and most of these are \blls. We cannot
say from these checks whether the unclassified \bcus~of the full catalogue
are mostly \blls, because the spectroscopic and reclassified samples favour
bright sources. The agreement with the classes is clearly better than chance
(AUC of 0.79 to 0.89 for \bcus), but the checks do not show that a given
tier is correct with the stated confidence, since the confidence is the
stability of a flag and not a probability (Section~\ref{sec:candidates_method}).

Five points limit the reading of these results. First, the spectroscopic
samples favour bright and accessible sources, which are not representative of
the \bcus~that stay unclassified, and the targeting criteria of each campaign
are not known. Second, a featureless spectrum is classified
as BL~Lac by default, so low signal-to-noise spectra make the BL~Lac labels
weaker than the FSRQ labels. Third, the hyperparameters were tuned on a random
80\% of all sources, which includes most of the DR1 and DR2 test sources, so
those two checks are not fully independent of the tuning. Fourth, the
SDSS-V classes are a proxy with an error of 12 to 16\%. Fifth, the samples
overlap and cannot be added.

\begin{deluxetable*}{lcccccc}
\tablecaption{
  Agreement of the single-class assignments with external classes. Sample is the source of the external classes; $n$ (\fsrq/\bll) is the number of sources, with the numbers of true \fsrqs~and \blls~in brackets; the next two columns give the number of correct calls over the number of calls of that class; All calls correct combines both classes; Coverage is the share of the sample that receives a single-class call; AUC is the area under the ROC curve of the difference between the \fsrq-mode and \bll-mode confidences. \label{tab:external}
}
\tablewidth{0pt}
\tablehead{
  \colhead{Sample} & \colhead{$n$ (\fsrq/\bll)} &
  \colhead{FSRQ-like calls correct} &
  \colhead{BL~Lac-like calls correct} &
  \colhead{All calls correct} &
  \colhead{Coverage} & \colhead{AUC}
}
\startdata
Published spectra (T1) & 42 (19/23)   & 15/18 (83\%) & 10/10 (100\%) & 25/28 (89\%)  & 67\% & 0.89 \\
SDSS-V proxy           & 89 (31/58)   & 20/26 (77\%) & 37/44 (84\%)  & 57/70 (81\%)  & 79\% & 0.79 \\
DR1 reclassified       & 236 (27/209) & 15/26 (58\%) & 170/178 (96\%) & 185/204 (91\%) & 86\% & 0.83 \\
DR2 reclassified       & 114 (15/99)  & 8/12 (67\%)  & 80/84 (95\%)  & 88/96 (92\%)  & 84\% & 0.83 \\
\enddata
\tablecomments{
  Calls are the \bcus~assigned to a single class at a confidence of 0.1.
  Coverage is the share of the sample that receives a single-class call.
  The 95\% Wilson intervals of the precision are 0.61 to 0.94 and 0.72 to
  1.00 (published spectra), 0.58 to 0.89 and 0.71 to 0.92 (SDSS-V), 0.39
  to 0.74 and 0.91 to 0.98 (DR1), and 0.39 to 0.86 and 0.88 to 0.98 (DR2)
  for FSRQ-like and BL~Lac-like calls. The samples overlap and are not
  added together.
}
\end{deluxetable*}

\subsection{SHAP Feature Importance and UMAP}
\label{sec:shap_results}
\label{sec:umap}

\begin{figure*}[ht]
\centering
\includegraphics[width=0.9\textwidth]{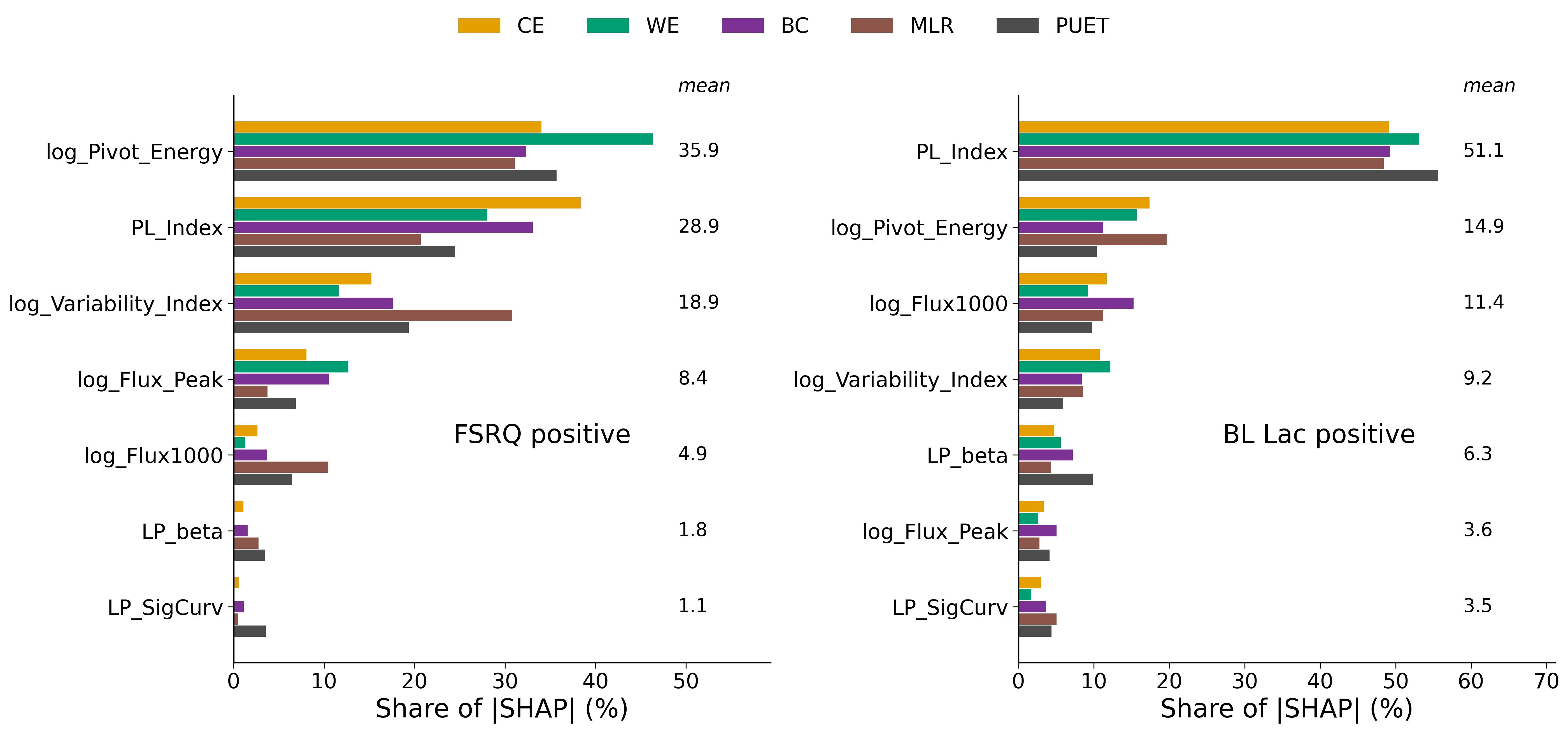}
\caption{Share of each feature in the total mean $|\mathrm{SHAP}|$ of each classifier (one bar colour per classifier), for \fsrq{} mode and \bll{} mode. The number at the right of each group is the mean over the five classifiers (Table~\ref{tab:shap}).}
\label{fig:shap}
\end{figure*}

Table~\ref{tab:shap} gives the share of each feature in the total
$|\mathrm{SHAP}|$ of a model, averaged over the five models, and
Figure~\ref{fig:shap} shows the share for each model.

\begin{deluxetable}{lcc}
\tablecaption{Share of each feature in the total mean $|\mathrm{SHAP}|$ (per cent), averaged over the five classifiers, for \fsrq{} mode and \bll{} mode. The shares of a mode add to 100 per cent.\label{tab:shap}}
\tablewidth{0pt}
\tablehead{
  \colhead{Feature} & \colhead{\fsrq{} mode} & \colhead{\bll{} mode}
}
\startdata
\texttt{log\_Pivot\_Energy} & 35.9 & 14.9 \\
\texttt{PL\_Index} & 28.9 & 51.1 \\
\texttt{log\_Variability\_Index} & 18.9 & 9.2 \\
\texttt{log\_Flux\_Peak} & 8.4 & 3.6 \\
\texttt{log\_Flux1000} & 4.9 & 11.4 \\
\texttt{LP\_beta} & 1.8 & 6.3 \\
\texttt{LP\_SigCurv} & 1.1 & 3.5 \\
\enddata
\tablecomments{
Shares are rounded to 0.1 and are computed before rounding, so a sum of
rounded entries can differ by 0.1 from the sum of the unrounded shares. For
example, the two curvature features in \bll{} mode are 6.33 and 3.54 before
rounding, which add to 9.9.
}
\end{deluxetable}

The spectral shape dominates in both modes. \texttt{PL\_Index} and
\texttt{log\_Pivot\_Energy} together carry 65\% of the total in \fsrq{} mode
and 66\% in \bll{} mode, but the split between them differs. In \bll{} mode
the spectral index alone carries 51\%, and in every classifier it carries
48--56\%. In \fsrq{} mode the pivot energy is first (35.9\%) and the index
second (28.9\%), and their order changes between classifiers: the index is
first in CE and BC, the pivot energy in WE, MLR and PUET. Because the two
features are strongly correlated, we do not read this order as a
statement about which one carries the physics.

The variability index is the third feature in \fsrq{} mode (18.9\%), and
MLR gives it 30.8\% against 11.6\% for WE. This agrees with the
higher fractional variability of \fsrqs~than of \blls~in 3-, 7- and 30-day
\fermi~light curves, where the mean for \bcus~lies between the two
classes \citep{Shah2025}. The two curvature features,
\texttt{LP\_beta} and \texttt{LP\_SigCurv}, carry 2.9\% together in \fsrq{} mode and
9.9\% in \bll{} mode. In the WE classifier for \fsrq{} mode their share is
0.0\%. The curvature is a minor input to the fitted models, most of all in \fsrq{} mode.

The classifiers differ in the scale of their scores, so the absolute
values differ as well. The summed $|\mathrm{SHAP}|$ per source is 0.04 for
BC in \fsrq{} mode against 0.37--0.41 for CE, WE and PUET, because its
scores lie between 0.34 and 0.63. Its shares in Table~\ref{tab:shap} are
nonetheless close to those of the other classifiers (32\% pivot energy, 33\%
index).

\begin{figure*}[ht]
\centering
\includegraphics[width=0.95\textwidth]{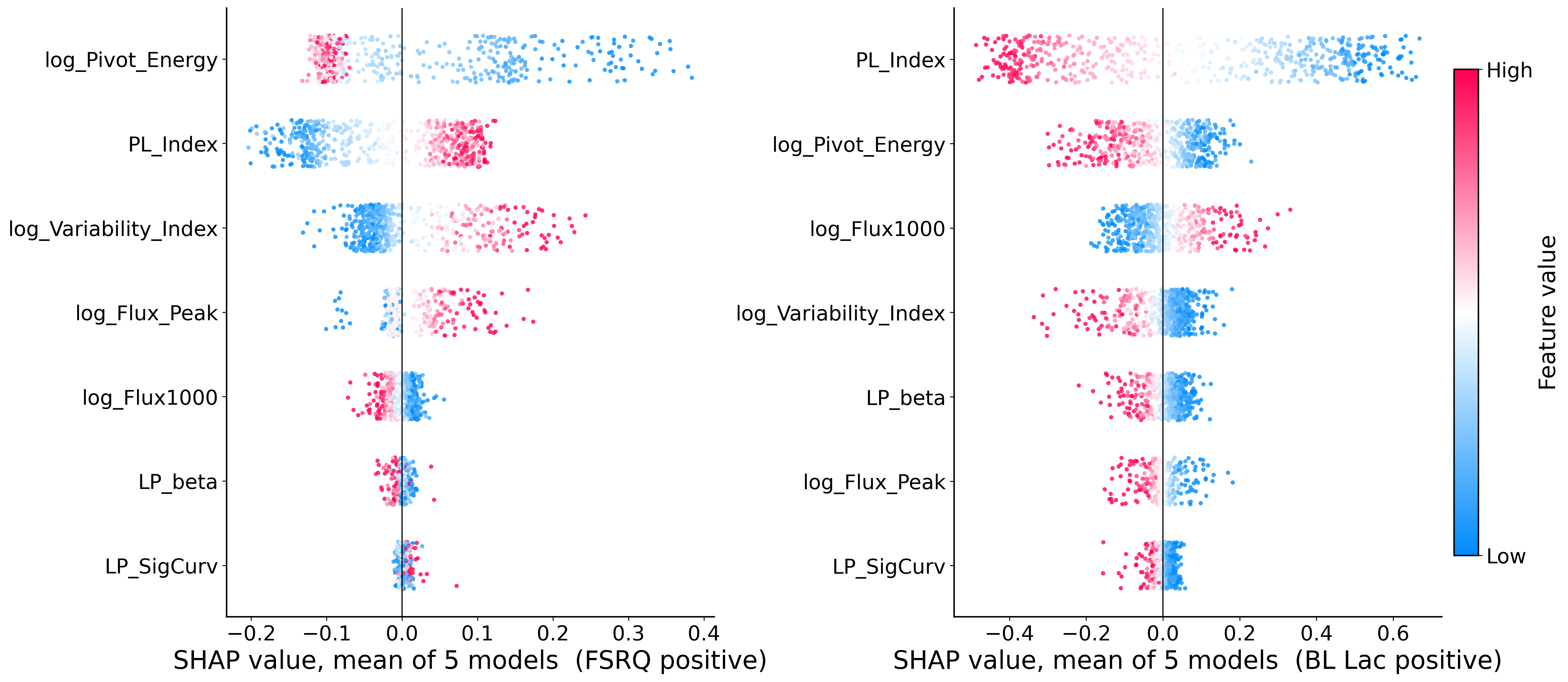}
\caption{SHAP values of the seven features for 600 sources, averaged over
the five classifiers, for \fsrq{} mode (left) and \bll{} mode (right).
Each point is one source. Its horizontal position is the SHAP value, the contribution of the feature to the score of the labelled class (positive values push the source towards that class), and its colour is the value of the feature, from low (blue) to high (red).}
\label{fig:beeswarm}
\end{figure*}

Figure~\ref{fig:beeswarm} shows the direction of the effects. In \fsrq{}
mode a source is pushed towards the FSRQ class by a soft spectrum (high
\texttt{PL\_Index}), a low pivot energy, a high variability index and a high
peak flux, and away from it by a high $1$--$100$~GeV flux. In \bll{} mode
the signs are the opposite for the index, the variability, the peak flux and
the $1$--$100$~GeV flux: a hard spectrum, a low variability, a low peak flux
and a high $1$--$100$~GeV flux favour the BL~Lac class. The pivot energy is
the exception: a low pivot energy raises the score in both modes, so it
does not separate the two classes by itself. The pivot energy and the
spectral index are strongly correlated (Section~\ref{sec:features}), so
the sign of its SHAP value need not follow the difference between the
classes in Figure~\ref{fig:feature_dist}; we do not interpret it further.
The two curvature features
have a small effect in both modes, and in \bll{} mode a high curvature
lowers the score.

\begin{figure}[ht]
\centering
\includegraphics[width=\columnwidth]{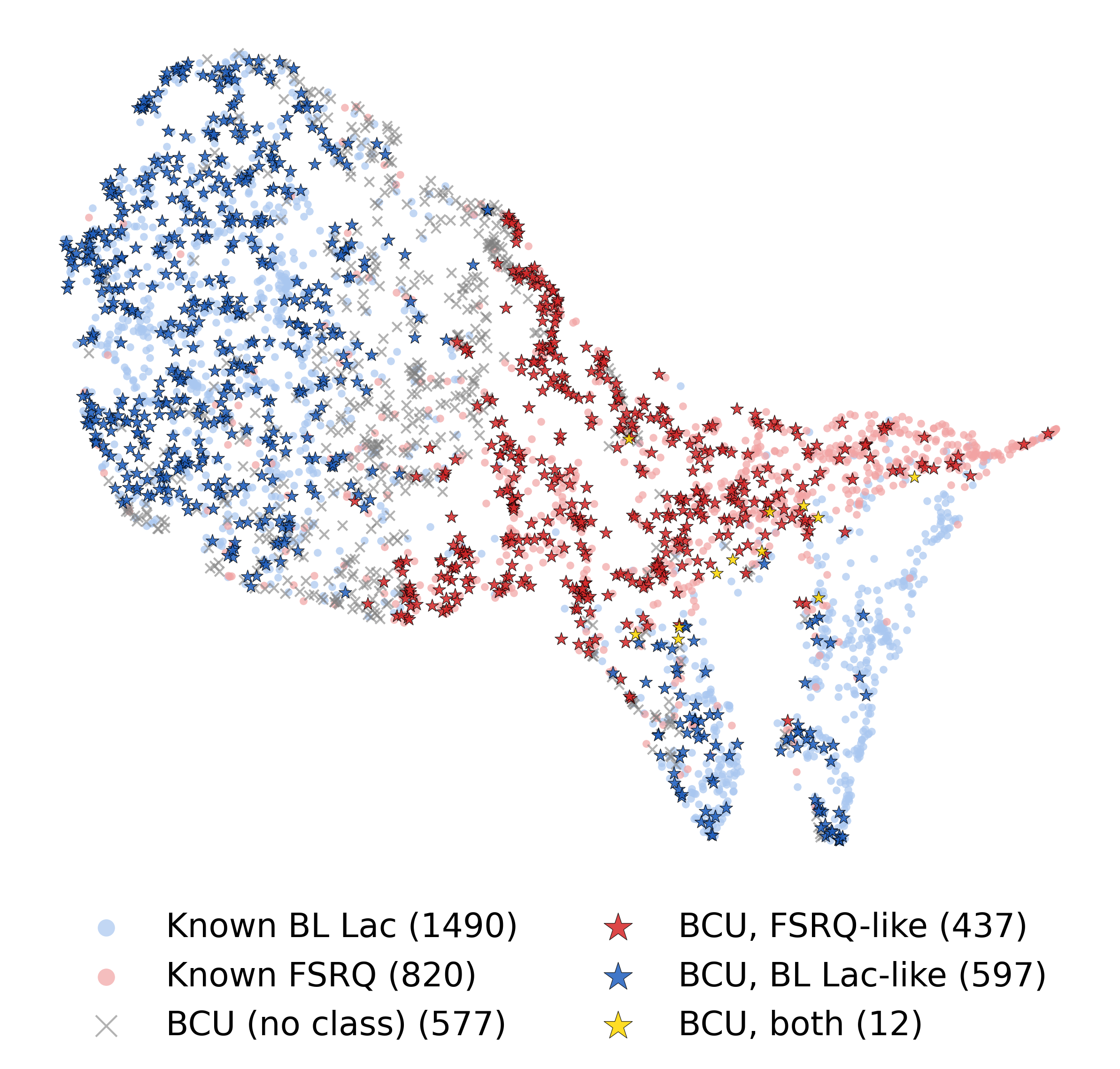}
\caption{UMAP projection of the seven standardised features. Known BL~Lacs
and FSRQs are shown as light blue and light red circles. The stars mark
BL~Lac-like (blue), FSRQ-like (red) and both-mode (yellow) candidates, and the crosses mark \bcus~without a class. The axes are the two UMAP coordinates and have no physical units.}
\label{fig:umap}
\end{figure}

Figure~\ref{fig:umap} shows a UMAP projection of all 3933 sources, computed
from the seven standardised features with 20 neighbours. UMAP only gives
a picture of the feature space and does not use the class labels, so it
does not test the classification. The known BL~Lacs fill a large region on
the left and extend into two branches at the bottom. The known FSRQs
form a band on the right, with a few lying inside the BL~Lac region. The
BL~Lac-like candidates sit in the BL~Lac region and the branches, and the
FSRQ-like candidates in the FSRQ band and along the arc that joins it to the
BL~Lac region. The \bcus~without a class are concentrated between the two
populations, with a smaller number inside the BL~Lac region. The twelve
sources flagged in both modes lie on the FSRQ side or in the lower branches,
and none lies in the dense part of the BL~Lac region.

\section{Discussion}
\label{sec:discussion}

\subsection{How Far the Classes Can Be Trusted}
\label{sec:trust}

The external checks give a consistent but uneven picture
(Table~\ref{tab:external}). BL~Lac-like calls agree with the external
class for 84 to 100\% of the sources, and for 95 to 96\% in the two largest
samples (DR1 and DR2). FSRQ-like calls agree for 58 to 83\%. In the DR1 and
DR2 tests the FSRQ-like precision is low partly because \blls~far outnumber
\fsrqs~among the sources that received a class. The recall for the
\fsrqs~of these groups (53 to 56\%) is also below that for the
known \fsrqs~(about 80\%). The headline recall of about 0.8 for known
\fsrqs~therefore overstates what to expect for FSRQ-like \bcus. A possible
reason is that \fsrqs~that remain \bcus~differ from the labelled \fsrqs~in
their gamma-ray features, but we did not test this.

The limits that apply to all the samples are listed in
Section~\ref{sec:external}. In addition, some DR4 labels are not spectroscopic.
\citet{Marais2024} observed southern 3FHL blazar candidates and note that
most of them are now classified as BL~Lac in the updated catalogues, but
that this does not appear to be a spectroscopic classification and rests,
for example, on the gamma-ray behaviour \citep{2016MNRAS.462.3180C}. Some
DR4 BL~Lac labels may therefore not be spectroscopic, and the
reclassification tests inherit that.

\subsection{PU Learning Against Supervised Baselines}
\label{sec:pu_vs_lr}

On the known sources the PU ensemble does not beat a logistic regression
trained on the same labels (Section~\ref{sec:baseline}). The logistic
regression gives $P(\mathrm{FSRQ})\geq 0.5$ for 97.9\% of the FSRQ-like
candidates and $P<0.5$ for all of the BL~Lac-like candidates. Of the \bcus~with
no class, 86.7\% have $0.1<P<0.9$, so the ensemble abstains mostly where
the logistic regression is uncertain. The two-mode scheme can leave a source without a class
or flag it in both modes, which a two-class model does not do on its own,
but we have not shown that this information is more accurate. We present
the ensemble as one way to produce class assignments with an abstention
option and not as a method shown to be more accurate than supervised
alternatives. The SCAR assumption behind the PU framing is unlikely to hold exactly here
(Section~\ref{sec:pu_framework}), which weakens the argument that PU
learning treats the \bcus~more correctly than a supervised classifier.

\subsection{Selection Effects and the Galactic Plane}
\label{sec:selection}

The labelled sources are a selected sample. The \bcus~are fainter, lie
closer to the plane and are newer in the catalogue
(Section~\ref{sec:data}). The scores of the \bcus~depend on whether the
sky coordinates are features (Section~\ref{sec:ablation}), although the
performance on known sources does not. A model that sees the coordinates
may learn where spectroscopy is available and carry that over to the
\bcus, so we leave them out. This removes about 2 of the 18 points by which
unclassified \bcus~are more common in the plane. The rest is not explained.
Fainter sources have noisier spectral parameters, and sources in the plane
are affected by the diffuse emission model, so both flux and position could
matter, and we did not separate them. A method that models the
probability that a source was labelled, as a function of flux and position,
and out-of-time validation on future releases would address this directly.
We leave both for future work.

\subsection{Sources Flagged in Both Modes}
\label{sec:transition}

Twelve \bcus~are candidates in both modes, and their gamma-ray features do
not separate the two classes. They could be transition objects between
\fsrqs~and \blls, or sources whose features are poorly measured
\citep{2011MNRAS.414.2674G,2013ApJ...764..135S}. Optical spectroscopy is
needed to decide. Table~\ref{tab:both} lists the twelve sources.

\begin{deluxetable}{lcc}
\tablecaption{\bcus~flagged in both modes. 4FGL is the source name without the prefix, and $c_{\rm FSRQ}$ and $c_{\rm BL}$ are the confidences in \fsrq{} mode and \bll{} mode (the fraction of the 100 repeats in which at least three of the five classifiers flag the source).\label{tab:both}}
\tablewidth{0pt}
\tablehead{
  \colhead{4FGL} & \colhead{$c_{\rm FSRQ}$} & \colhead{$c_{\rm BL}$}
}
\startdata
J0204.8+1513 & 1.00 & 0.41 \\
J0207.5$-$2402 & 0.96 & 0.98 \\
J0216.6$-$1015 & 0.81 & 0.53 \\
J0223.5$-$0928 & 0.96 & 0.50 \\
J0347.7$-$3616 & 0.11 & 0.67 \\
J0418.1$-$0252 & 0.99 & 0.28 \\
J0420.3$-$3745 & 0.63 & 1.00 \\
J1339.0$-$2400 & 1.00 & 1.00 \\
J1416.0+1320 & 1.00 & 1.00 \\
J1421.1$-$1120 & 1.00 & 0.71 \\
J2104.0$-$3546 & 1.00 & 0.13 \\
J2325.7+1821 & 0.57 & 0.29 \\
\enddata
\end{deluxetable}

\subsection{Comparison with Previous Work}
\label{sec:comparison}

\citet{2023ApJ...946..109A} classified the 1115 \bcus~of the 4LAC-DR3 Clean
sample with five supervised classifiers and labelled a \bcu~only if all
five agreed, which gave 610 \blls~and 333 \fsrqs~(943 sources, 85\%) and an
accuracy of $0.938\pm0.012$ on known sources. We assign a single class to
1034 of 1623 \bcus~(64\%). The catalogues and the feature sets differ, so
the counts are not directly comparable. The accuracy of
\citet{2023ApJ...946..109A} is measured on known sources. The corresponding
number for our ensemble is a recall of about 0.8 at a false-positive rate of
0.08. On \bcus~that were later classified, our BL~Lac-like calls agree
with the external class for 95 to 96\% of the sources in DR1 and DR2 and our
FSRQ-like calls for 58 to 67\%. These are per-class agreement rates on a
selected sample, not an accuracy over both classes, so they cannot be set
against 0.938 directly. As in their study, accuracy on known sources does not
measure accuracy on \bcus.

\subsection{Limitations and Future Work}
\label{sec:limitations}

The main limitation is the restriction to gamma-ray features. The
hyperparameters were tuned with nine features on a random 80\% of the
sources, and were not retuned for the seven-feature
ensemble. The external samples are small and selected, and the FSRQ-like
precision rests on 12 to 26 sources per sample. \texttt{log\_Flux\_Peak} is
missing for 57.7\% of the sources, and we did not test its removal. The
final run uses 100 repeats, against 200 for the nine-feature run. The
excess of unclassified \bcus~in the Galactic plane is not explained
(Section~\ref{sec:selection}).

Future work should model the selection of labelled sources explicitly,
for example with selection-aware positive-unlabelled methods and
validation on sources classified after the training catalogue, and should
add multiwavelength information to the gamma-ray features.

\section{Summary and Conclusions}
\label{sec:conclusions}

We applied an ensemble of five positive-unlabelled classifiers to the
1623 \bcus~in \fermi~\fourFGL-DR4, using seven gamma-ray features and
out-of-fold scoring. Our main findings are as follows.

\begin{enumerate}

\item At a confidence of at least 0.1, 437 \bcus~are FSRQ-like and 597
are BL~Lac-like, 12 are candidates in both modes and 577 in neither.
Of the single-class candidates, 829 (80\%) have a confidence of at least
0.9.

\item On known sources the ensemble has a recall of 0.80 (\fsrq{} mode) and
0.82 (\bll{} mode) at a false-positive rate of 0.08. A logistic regression
trained on the labelled sources alone reaches a recall of 0.85 at a
false-positive rate of 0.077, so the PU framing gives no accuracy gain on
known sources.

\item Removing the Galactic coordinates does not lower the performance on
known sources (the vote recall changes by $+0.013$ and $-0.007$), but they change the \bcu~scores: removing them adds 107 \bcus~to
the candidate lists, 56\% of them in tier C, and drops 30. We exclude them. The excess of
unclassified \bcus~in the Galactic plane (16 percentage points after the
removal) is not explained.

\item Against published spectra, SDSS-V proxy classes, and \bcus~that
later received a class, BL~Lac-like calls are right for 84 to 100\% of
the sources and FSRQ-like calls for 58 to 83\%. The \fsrqs~among the
reclassified \bcus~are found less often (53 to 56\%) than known \fsrqs~(about 80\%).
These samples are selected and overlap.

\item The \bcus~are fainter and closer to the plane than the labelled
sources, so the labelled sample does not represent them. Accuracy on known
sources should not be read as accuracy on \bcus.

\item A table with the confidences and tiers of all 1623 \bcus~is made
available at Zenodo (\url{https://doi.org/10.5281/zenodo.23609167}).

\end{enumerate}

\begin{acknowledgments}

S.A.\ and Z.S.\ thank the Inter-University Centre for Astronomy and
Astrophysics (IUCAA), Pune, India, for support and facilities. This work uses
data from the \textit{Fermi} Large Area Telescope; we thank the \textit{Fermi}-LAT
Collaboration for the public 4FGL-DR4 catalogue.

\end{acknowledgments}

\vspace{5mm}

\software{
  \texttt{numpy} \citep{Harris2020},
  \texttt{pandas} \citep{McKinney2010},
  \texttt{scikit-learn} \citep{2011JMLR...12.2825P},
  \texttt{optuna} \citep{2019KDD....14.2623A},
  \texttt{pulearn} \citep{pulearn},
  \texttt{shap} \citep{Lundberg2017},
  \texttt{umap-learn} \citep{McInnes2018},
  \texttt{astropy} \citep{2022ApJ...935..167A},
  \texttt{matplotlib} \citep{Hunter2007},
  \texttt{scipy} \citep{Virtanen2020},
  \texttt{xgboost} \citep{Chen2016},
  \texttt{lightgbm} \citep{Ke2017},
  \texttt{imbalanced-learn} \citep{Lemaitre2017}
}

\appendix

\section{Base-Classifier Selection}
\label{app:supervised}

Table~\ref{tab:supervised} lists the eight supervised classifiers that were
tuned on the labelled FSRQs and BL~Lacs. Recall and $L_2$ are means over
1000 random 80/20 splits of the tuning pool. The three classifiers with the highest recall in
each mode (bold) were the candidates for the base classifier of CE, WE and
BC, following the recall rule of Section~\ref{sec:classifiers}. Ranking by
$L_2$ instead of recall would select different sets, namely random forest,
XGBoost and LightGBM for FSRQs, and logistic regression, XGBoost and
LightGBM for BL~Lacs.

\begin{deluxetable}{lcccc}
\tablecaption{Supervised classifiers tuned for each mode (\fsrq{} or \bll{} is the labelled positive class). Recall is the held-out recall, as mean$\pm$standard deviation over 1000 random 80/20 splits of the tuning pool, and $L_2$ is the mean Lee--Liu score over the same splits. Bold marks the three highest recalls. \label{tab:supervised}}
\tablewidth{0pt}
\tablehead{
  \colhead{} & \multicolumn{2}{c}{FSRQ} & \multicolumn{2}{c}{BL~Lac} \\
  \colhead{Classifier} & \colhead{Recall} & \colhead{$L_2$} & \colhead{Recall} & \colhead{$L_2$}
}
\startdata
AdaBoost & \textbf{0.837}$\pm$0.032 & 2.12 & 0.810$\pm$0.025 & 1.46 \\
LDA & \textbf{0.831}$\pm$0.031 & 2.07 & \textbf{0.825}$\pm$0.023 & 1.49 \\
Logistic regression & \textbf{0.823}$\pm$0.032 & 2.09 & \textbf{0.817}$\pm$0.024 & 1.50 \\
XGBoost & 0.822$\pm$0.033 & 2.16 & 0.780$\pm$0.026 & 1.50 \\
SVC & 0.813$\pm$0.033 & 2.09 & \textbf{0.812}$\pm$0.025 & 1.48 \\
LightGBM & 0.785$\pm$0.036 & 2.15 & 0.779$\pm$0.027 & 1.50 \\
Random forest & 0.782$\pm$0.035 & 2.16 & 0.770$\pm$0.027 & 1.46 \\
Decision tree & 0.760$\pm$0.044 & 1.94 & 0.737$\pm$0.035 & 1.32 \\
\enddata
\end{deluxetable}

\section{Candidate Table}
\label{app:candidates}

Table~\ref{tab:cand_sample} shows a few rows of the released table of confidences and tiers (Zenodo, \url{https://doi.org/10.5281/zenodo.23609167}; file \texttt{bcu\_classification\_all.csv}). The file has one row for each of the 1623 \bcus: the 1046 candidates (437 FSRQ-like, 597 BL~Lac-like and 12 ambiguous) and the 577 \bcus~without a class. Its columns are the \fourFGL~source name, the assignment, the tier, the confidence in each mode, the confidence of each of the five classifiers in each mode, $n_{\rm agree}$, and the seven feature values. The per-classifier confidences are given for the candidates only, and $n_{\rm agree}$ for the single-class candidates only. Confidence is the fraction of repeats in which the source is flagged (Section~\ref{sec:candidates_method}), and $n_{\rm agree}$ is the number of the five classifiers whose own confidence in the assigned class is at least 0.5. It can be smaller than 3 for a source in tier C, because the confidence of the ensemble counts the repeats in which three classifiers flag the source together, not the number of classifiers that flag it in most repeats. The rows in Table~\ref{tab:cand_sample} are the first sources by name in each tier, plus two sources flagged in both modes.

\begin{deluxetable}{llcccc}
\tablecaption{Sample rows of the released table. 4FGL is the source name without the prefix; Assignment is the class assigned at a confidence of 0.1 (ambiguous: a candidate in both modes); Tier is the stability tier of the assigned class (none for ambiguous sources); $c_{\rm FSRQ}$ and $c_{\rm BL}$ are the confidences in \fsrq{} mode and \bll{} mode; $n_{\rm agree}$ is the number of the five classifiers whose own confidence in the assigned class is at least 0.5.\label{tab:cand_sample}}
\tablewidth{0pt}
\tablehead{
  \colhead{4FGL} & \colhead{Assignment} & \colhead{Tier} & \colhead{$c_{\rm FSRQ}$} & \colhead{$c_{\rm BL}$} & \colhead{$n_{\rm agree}$}
}
\startdata
J0001.2+4741 & BL~Lac-like & A & 0.00 & 0.94 & 4 \\
J0001.6$-$4156 & BL~Lac-like & A & 0.00 & 1.00 & 5 \\
J0001.8$-$2153 & BL~Lac-like & A & 0.00 & 0.95 & 3 \\
J0003.1$-$5248 & BL~Lac-like & A & 0.00 & 1.00 & 5 \\
J0002.4$-$5156 & BL~Lac-like & B & 0.00 & 0.61 & 3 \\
J0007.5$-$5247 & FSRQ-like & B & 0.78 & 0.00 & 3 \\
J0028.1+7505 & BL~Lac-like & B & 0.00 & 0.55 & 3 \\
J0003.3$-$1928 & FSRQ-like & C & 0.43 & 0.05 & 3 \\
J0019.2$-$5640 & FSRQ-like & C & 0.31 & 0.01 & 2 \\
J0035.8+6131 & FSRQ-like & C & 0.42 & 0.00 & 3 \\
J0204.8+1513 & ambiguous & -- & 1.00 & 0.41 & -- \\
J0207.5$-$2402 & ambiguous & -- & 0.96 & 0.98 & -- \\
\enddata
\end{deluxetable}

\section{Optimised PU Classifier Hyperparameters}
\label{app:hyperparams}

Table~\ref{tab:hyperparams} gives the hyperparameters found by the tuning
of Section~\ref{sec:hyperopt} (150 trials per classifier, base classifiers
chosen by the recall rule). They were tuned with the nine-feature set and
are used unchanged for the seven-feature analysis. The last column is the
cross-validated $L_2$ of the tuning. The LDA parameters are the scikit-learn
settings listed.

\begin{deluxetable*}{llllc}
\tabletypesize{\footnotesize}
\tablecaption{Hyperparameters of the PU classifiers. Mode is the labelled positive class, PU method is the classifier (Section~\ref{sec:classifiers}), Base classifier is the supervised model used inside it, Other parameters are the remaining tuned values, and cv $L_2$ is the cross-validated Lee--Liu score of the tuning. BC bags the base
classifier named in the third column, with the number of estimators (est.), the fraction of sources (samples) and the fraction of features (features) drawn for each. The PUET parameters are the risk
estimator, the loss, the number of trees, the minimum leaf size, the number
of features per split and \texttt{max\_candidates}.
  \label{tab:hyperparams}}
\tablewidth{0pt}
\tablehead{
  \colhead{Mode} & \colhead{PU method} & \colhead{Base classifier} & \colhead{Other parameters} & \colhead{cv $L_2$}
}
\startdata
FSRQ & CE & AdaBoost, $n=92$, $\eta=0.1234$ & hold-out ratio 0.105 & 2.200 \\
FSRQ & WE & AdaBoost, $n=92$, $\eta=0.1234$ & hold-out ratio 0.290 & 2.197 \\
FSRQ & BC & AdaBoost, $n=92$, $\eta=0.1234$ & 67 est., samples 0.298, features 0.619 & 2.191 \\
FSRQ & MLR & -- & learning rate 0.0856 & 1.953 \\
FSRQ & PUET & -- & uPU, quadratic, 170 trees, leaf 8, sqrt, 2 & 2.251 \\
\hline
BL~Lac & CE & LDA, lsqr & hold-out ratio 0.142 & 1.495 \\
BL~Lac & WE & LDA, lsqr & hold-out ratio 0.143 & 1.462 \\
BL~Lac & BC & logistic reg., $C=0.147$, lbfgs & 119 est., samples 0.692, features 0.980 & 1.492 \\
BL~Lac & MLR & -- & learning rate 0.0654 & 1.474 \\
BL~Lac & PUET & -- & uPU, quadratic, 154 trees, leaf 9, sqrt, 2 & 1.505 \\
\enddata
\tablecomments{$\eta$ is the AdaBoost learning rate and $n$ the number of
estimators. The logistic regression has max\_iter 227 and tolerance
$1.96\times10^{-4}$.}
\end{deluxetable*}

\bibliography{bcu_paper}{}
\bibliographystyle{aasjournal}

\end{document}